\documentclass[12pt]{article}
\pdfoutput=1

\usepackage[bulletsep]{collref}
\usepackage{amssymb,graphicx}
\usepackage[intlimits]{amsmath}
\usepackage{pst-all}
\usepackage{bbm}

\makeatletter \@addtoreset{equation}{section} \makeatother

\makeatletter
\let\old@startsection=\@startsection
\let\oldl@section=\l@section
\renewcommand{\@startsection}[6]{\old@startsection{#1}{#2}{#3}{#4}{#5}{#6\mathversion{bold}}}
\renewcommand{\l@section}[2]{\oldl@section{\mathversion{bold}#1}{#2}}
\makeatother

\newcommand{\strokedint}{-\!\!\!\!\!\!\int}
\newcommand{\strokedintmore}{-\!\!\!\!\!\!\!\int}

\begin{document}

\begin{flushright}\footnotesize
\texttt{HU-EP-11/27}\\
\texttt{ITEP-TH-18/11}\\
\texttt{NORDITA-2011-48} \\
\texttt{UUITP-19/11}
\vspace{0.4cm}
\end{flushright}

\renewcommand{\thefootnote}{\fnsymbol{footnote}}
\setcounter{footnote}{0}

\begin{center}
{\Large\textbf{\mathversion{bold} Wilson Loops in $N=2$ Super-Yang-Mills \\
from Matrix Model}
\par}

\vspace{0.8cm}

\textrm{F.~Passerini$^{1}$ and
K.~Zarembo$^{2,3}$\footnote{Also at ITEP, Moscow, Russia}}
\vspace{4mm}

\textit{${}^1$Institut f\"ur Physik, Humboldt-Universit\"at zu Berlin\\
Newtonstrasse 15, D-12489 Berlin, Germany}\\
\textit{${}^2$Nordita,
Roslagstullsbacken 23, SE-106 91 Stockholm, Sweden}\\
\textit{${}^3$Department of Physics and Astronomy, Uppsala University\\
SE-751 08 Uppsala, Sweden}\\
\vspace{0.2cm}
\texttt{filippo@physik.hu-berlin.de, zarembo@nordita.org}

\vspace{3mm}


\par\vspace{1cm}

\textbf{Abstract} \vspace{3mm}

\begin{minipage}{13cm}
We compute the expectation value of the circular Wilson loop in $\mathcal{N}=2$ supersymmetric Yang-Mills theory with $N_f=2N$ hypermultiplets. Our results indicate that the string tension in the dual string theory scales as the logarithm of the 't~Hooft coupling.
\end{minipage}

\end{center}

\vspace{0.5cm}


\newpage
\setcounter{page}{1}
\renewcommand{\thefootnote}{\arabic{footnote}}
\setcounter{footnote}{0}

\section{Introduction and Summary}

Exact results in quantum field theory usually rely on powerful symmetry principles, such as supersymmetry. A beautiful example of non-perturbative use of supersymmetry is Pestun's exact calculation of circular Wilson loops in a wide class of $\mathcal{N}=2$ supersymmetric Yang-Mills theories \cite{Pestun:2007rz}. To be more precise, Pestun reduced the problem to a finite-dimensional matrix integral, still to be evaluated. In this paper we address this problem in the 't~Hooft large-$N$ limit.

The large-$N$ limit of Pestun's matrix model has been discussed in \cite{Rey:2010ry}, with rather unexpected conclusions on its strong-coupling behavior. The conclusions of \cite{Rey:2010ry} for the most part rely on simple scaling arguments. Our goal will be to develop systematic strong and weak coupling expansions of  Pestun's matrix model and its observables. We shall see that the strong-coupling behavior of the matrix integral is indeed very unusual. The large-$N$ master field has an infinite support at infinite coupling, but such that the majority of matrix eigenvalues remain finite in accord with the scaling arguments.  The limiting master field describes a certain class of observables, but not all observables. In particular, it does not describe Wilson loops, which are determined by the largest eigenvalue as in ordinary matrix models.

We shall concentrate on the $\mathcal{N}=2$ superconformal Yang-Mills theory (SCYM), an $SU(N)$ gauge theory with $N_f=2N$ hypermultiplets in the fundamental representation. This theory has zero beta function and therefore is superconformal at any value of the Yang-Mills coupling $g$. One may expect that at $N\rightarrow \infty $ and when the 't~Hooft coupling $\lambda =g^2N$ is large the SCYM theory is described by a weakly-coupled dual string theory. There is a number of proposals for the dual string/supergravity background \cite{Gaiotto:2009gz,Gadde:2009dj,ReidEdwards:2010qs}. 
Neither of these backgrounds is of the simple form $AdS_5\times X^5$. In \cite{Gaiotto:2009gz,ReidEdwards:2010qs}, the $AdS_5$ part of the type-IIA/M-theory geometry is warped with respect to the coordinates of $X^5$. In \cite{Gadde:2009dj}, the string dual is non-critical, so $X^5$ is less than five-dimensional, or even partly non-geometric. In either case the dual string theory is never completely classical, since the backgrounds of \cite{Gaiotto:2009gz,ReidEdwards:2010qs}  contain a curvature singularity, and the non-critical string of  \cite{Gadde:2009dj}  is intrinsically quantum.

We believe that Wilson loop calculations (along with possible integrability of the SCYM theory \cite{Gadde:2010zi,Gadde:2010ku,Pomoni:2011jj,Liendo:2011xb}) may shed more light on the AdS/CFT duality in the $\mathcal{N}=2$ setting.
For one thing, Wilson loops are very sensitive probes of the dual string dynamics, as they couple directly to the string worldsheet \cite{Maldacena:1998im,Rey:1998ik}. One can also do exact field-theory calculation at any coupling with the help of Pestun's results \cite{Pestun:2007rz}. In the more familiar $\mathcal{N}=4$
case, the exact calculation of the circular Wilson loop \cite{Erickson:2000af,Drukker:2000rr,Pestun:2007rz} immediately confirms the AdS/CFT relationship between the string tension and the 't~Hooft coupling \cite{Maldacena:1998re}\footnote{By the string tension we will always mean the dimensionless ratio of the AdS radius squared to $\alpha'$: $T=R^2/2\pi\alpha'$.}:
\begin{equation}\label{N=4}
 T_{\mathcal{N}=4}=\frac{\sqrt{\lambda_{\mathcal{N}=4}}}{2\pi}\,.
\end{equation}
Our goal will be to derive a similar relationship for $\mathcal{N}=2$ SCYM.

The circular Wilson loops in $\mathcal{N}=4$ SYM were first calculated by resumming planar diagrams \cite{Erickson:2000af,Drukker:2000rr}. The planar perturbation theory for Wilson loops in $\mathcal{N}=2$ SCYM has been studied in \cite{Andree:2010na}, where quite interesting regularities have been observed. To make connection to this work we will also study the weak-coupling expansion of Pestun's matrix model.

The paper is organized as follows: in the next subsection we briefly summarize our strong-coupling results and discuss their possible implication for the string dual of $\mathcal{N}=2$ SCYM. In sec.~\ref{sec:partition} we review Pestun's matrix model for $\mathcal{N}=2$ SCYM on $S^4$ and derive the saddle-point equations for the eigenvalue distribution in its large-$N$ limit. In sec.~\ref{sec:weak} we study the weak and in sec.~\ref{sec:strong} the strong coupling expansions of the saddle-point equations, and in sec.~\ref{sec:instantons} we compute the one-instanton correction. We conclude with general discussion in sec.~\ref{sec:conclusions}. The technical details of our calculations are collected in the appendices.

\subsection{Summary of results}

The vacuum expectation value of the Wilson loop is computed holographically by summing over random surfaces in the bulk which end on the given contour on the boundary \cite{Maldacena:1998im,Rey:1998ik}. Assuming that the string tension is large, the string path integral is saturated by the surface of the minimal area. For the circular loop on the boundary of $AdS_5$ the regularized minimal area (which is negative for any contour) is equal to $-2\pi $ \cite{Drukker:1999zq,Berenstein:1998ij}.  The expected form of the circular Wilson loop expectation value in the semiclassical regime then has the form
\begin{equation}\label{WT}
 W(C_{\rm circle})=K\,\cdot T^{-3/2}\,{\rm e}\,^{2\pi T} \qquad
 (T\rightarrow \infty ),
\end{equation}
where $T$ is the string tension. The factor of $T^{-3/2}$ comes from the gauge fixing in the string path integral, as explained in \cite{Drukker:2000rr}, and the constant factor $K$ is determined by the  quantum fluctuations of the string worldsheet.

The semiclassical calculation of the circular Wilson loop was sensitive only to the universal $AdS_5$ factor in the geometry, that has to be there because of the conformal symmetry. The structure of the internal space ($X^5$) was not very important, and could only enter through the prefactor in the formula (\ref{WT}). In this respect the result (\ref{WT}) looks completely universal and should apply to any theory with the AdS dual. However, the derivation also assumes that the semiclassical approximation is accurate in some range of parameters, which might or might not be the case. It might happen that the string dual of $\mathcal{N}=2$ SCYM is always in the quantum regime, either because of the curvature singularities present in the background geometry \cite{Gaiotto:2009gz,ReidEdwards:2010qs}, or because of strong quantum fluctuations that cancel the  central charge in case the string dual is non-critical \cite{Gadde:2009dj}. Whether or not the quantum sector in the string sigma-model can be separated from the geometric $AdS_5$ factor in the Wilson loop calculation, and whether the curvature of $AdS_5$ ever becomes small or not is unclear to us. In some sense (\ref{WT}) can be regarded as a parameterization of the Wilson loop vev in terms of the effective string tension. We will find that the circular Wilson loop vev in $\mathcal{N}=2$ SCYM is indeed consistent with the parameterization (\ref{WT}) upon a simple identification
\begin{equation}\label{stringtension}
 T=\frac{3}{2\pi }\,\ln\lambda .
\end{equation}
This is certainly very different from the standard $\mathcal{N}=4$ relation (\ref{N=4}).

The logarithmic behavior of the effective string tension follows from the power-like growth of the Wilson loop expectation value at strong coupling:
\begin{equation}\label{llambdasca}
 W(C_{\rm circle})=\,{\rm const}\,\,\frac{\lambda ^3}{\left(\ln\lambda\right)^{3/2} }\,.
\end{equation}
We will try to be accurate with normalization, although keeping constants on top of the logarithms is always  difficult. Our estimate for the coefficient $K$ in (\ref{WT})  is
\begin{equation}\label{Kconst}
 K\simeq 3.13\cdot 10^{-5}.
\end{equation}
The corresponding constant in (\ref{llambdasca}) is $9.47\cdot 10^{-5}$. These are analytic but approximate predictions, which we expect to have a few percent accuracy.

\section{Partition function and saddle-point equations}\label{sec:partition}

The field content of the $\mathcal{N}=2$ SCYM theory consists of the $SU(N)$ gauge field $A_\mu $, two adjoint scalars $\Phi_I $ from the vector multiplet ($I=1,2$), $2\times 2N$ fundamental scalars $Q^A_f$, $\bar{Q}^f_A$  from $2N$ hypermultiplets ($A=1,2$; $f=1,\ldots, 2N$), and various fermions that make the spectrum supersymmetric. The Wilson loop is defined as
\begin{equation}
 W(C)=\left\langle \frac{1}{N}\,\mathop{\mathrm{tr}}
 {\rm P}\exp\left[\int_{C}^{}ds\,\left(iA_\mu (x)\acute{x}^\mu +n_I\Phi_I (x)|\acute{x}|\right)\right]\right\rangle,
\end{equation}
where $\mathbf{n}$ is a unit two-dimensional vector, which can vary along the contour, but  in Pestun's calculation has fixed constant components.

\subsection{Partition function on $S^4$}

Pestun computed the partition function of $\mathcal{N}=2$ SCYM on $S^4$  using localization \cite{Pestun:2007rz}. Compactification on the sphere provides a useful IR regularization, but otherwise is not important, since we are dealing with a conformal theory and the sphere is conformally equivalent to $\mathbb{R}^4$. Let us briefly review the main steps of Pestun's calculation. 

One can define a ``vacuum state" of $\mathcal{N}=4$ SCYM ``on the Coulomb branch" by Higgsing the theory with an expectation value of the scalar field $\Phi _I$ along the $\mathbf{n}$ direction. The expectation value can be brought to the diagonal form, with respect to the $SU(N)$ indices, by a gauge transformation:  $\left\langle \Phi _I\right\rangle=n_I\mathop{\mathrm{diag}}(a_1,\ldots ,a_N)$. The eigenvalues $a_i$ should satisfy the $SU(N)$ constraint:
\begin{equation}\label{traceless}
 \sum_{i=1}^{N}a_i=0.
\end{equation}
On the sphere one actually has to average over all vacua, but this can be postponed till the last moment. A useful strategy is to first compute an effective action for $a_i$ by integrating out all other fields. Because of the supersymmetry this effective action is one-loop exact. A proof and the explicit calculation of the one-loop factors can be found in \cite{Pestun:2007rz}.

The path integral of $\mathcal{N}=2$ SCYM on $S^4$ thus reduces to a matrix integral over the zero mode of the adjoint scalar\footnote{We choose to normalize the Yang-Mills kinetic term as $ -\mathop{\mathrm{tr}}F_{\mu \nu }^2/2g^2$, which differs by a factor of $2$ from the conventions in \cite{Pestun:2007rz}.}:
\begin{equation}\label{Pestunmatrix}
 Z= \int d^{N-1}a \,\prod_{i<j}^{}\left(a_i-a_j\right)^2\,{\rm e}\,^{-\frac { 8\pi^2 } {g^2}\,\sum\limits_{i} a_i ^2 } {\cal Z}_{\text{1-loop}}(a)\left|{\cal Z }_\text{inst}(a;g^2)\right|^2
\end{equation}
The first factor in the integrand (the Vandermonde determinant) is the Faddeev-Popov determinant of the diagonal gauge.
The exponent in the second factor is the classical action -- the area of the four-sphere times the conformal coupling of the scalar to the constant curvature of $S^4$. The last two terms are the one-loop determinant from integrating out field fluctuations and the instanton contribution.

The instanton partition function $\mathcal{Z}_{\rm inst}(a;g^2)$ is a known 
\cite{Nekrasov:2002qd,Nekrasov:2003rj}, albeit fairly complicated function of the eigenvalues $a_i$ and the Yang-Mills coupling. It is usually assumed that instantons are not important in the large-$N$ limit, because the instanton action is linear in $N$ 
in the large-$N$ limit:
$$
\,{\rm e}\,^{-\frac{8\pi^2}{g^2}}=\,{\rm e}\,^{-\frac{8\pi^2}{\lambda}\,N}.
$$
In practice the situation is more complicated, because the instanton weight contains moduli integration and the number of instanton moduli grows with $N$. Whether instantons are suppressed or not is thus a dynamical question \cite{Gross:1994mr}. There are known examples where the volume of the moduli space overcomes the suppression by the instanton action and leads to a large-$N$ phase transition into a non-perturbative phase. Later we will compute the one-instanton weight to check if there is an instanton-driven phase transition in $\mathcal{N}=2$ SCYM. We will find that the moduli integration enhances the instanton weight by a factor of $\sqrt{N}$, which is insufficient to overcome the exponential suppression by the instanton action. 
For now on we  just set $\mathcal{Z}_{\rm inst}(a;g^2)=1$.

The scalar vev acts on the adjoint fields from the vector multiplet through the commutator: $[\left\langle \Phi\right\rangle,v]_{ij}=(a_i-a_j)v_{ij}$ and therefore the vector multiplet fields get masses  $m_{ij}^2\sim(a_i-a_j)^2$. On the fundamental fields in the hypermultiplets the scalar vev acts by multiplication: $(\left\langle \Phi\right\rangle h)_i= a_ih_i$ and the hypermultiplets get masses $m_i^2\sim a_i^2$. The one-loop effective action is thus a combination of terms that depend on $(a_i-a_j)^2$ and $a_i^2$. The explicit expression was computed by Pestun, and can be expressed through a single function $H(z)$, which is  related to the Barnes $G$-function:
$$H(x)=\,{\rm e}\,^{-(1+\gamma) x^2}G(1+ix)G(1-ix),$$ 
and admits an infinite-product representation:
\begin{equation}\label{def:H}
  H(x) = \prod_{n=1}^{\infty} \left[
   \left(1 +\frac {x^2}{n^2}\right)^{n}
 \,{\rm e}\,^{-\frac{x^2} {n}}\right].
\end{equation}
The one-loop factor in the partition function is given by \cite{Pestun:2007rz}
\begin{equation}\label{1loop}
 {\cal Z}_{\text{1-loop}}=\frac{\prod\nolimits_{ i <j} H^2(a_i-a_j)}{\prod\nolimits_{i} H^{2N}(a_i)},
\end{equation}
where the numerator  is the contribution of the vector multiplet and the denominator is the contribution of $2N$ hypermultiplets. 

\begin{figure}[t]
\centerline{\includegraphics[width=2cm]{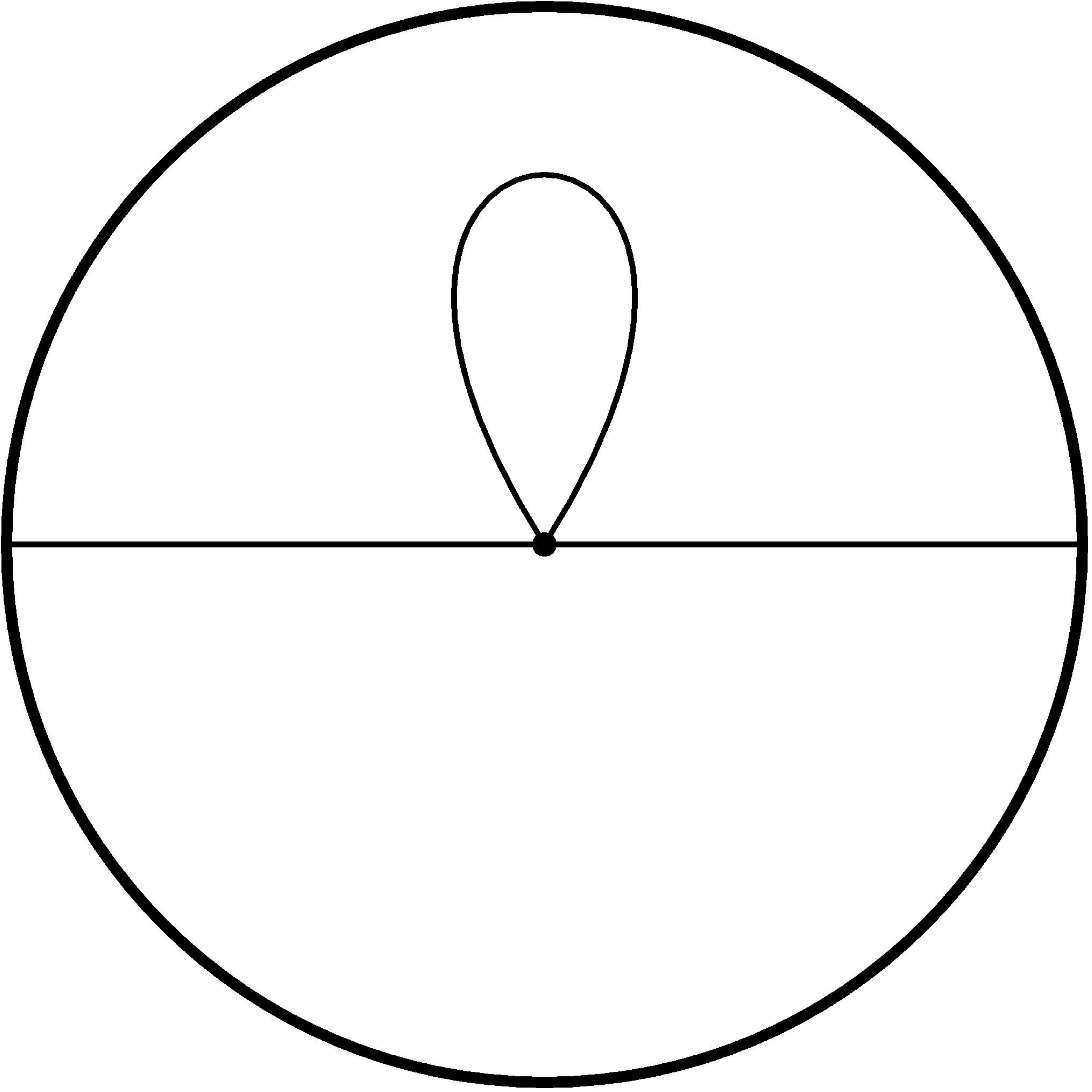}}
\caption{\label{fig:diagrams}\small The lowest-order non-Gaussian correction to the Wilson loop in the matrix model appears at order $O(\lambda^3)$.}
\end{figure}

The expression (\ref{1loop}) has an interesting symmetry. If we multiply $H(z)$ by a Gaussian:
\begin{equation}\label{shiftsymm}
 H(z)\rightarrow H(z)\,{\rm e}\,^{Cz^2},
\end{equation}
with an arbitrary constant $C$, the partition function (\ref{1loop}) will not change because of the trace condition (\ref{traceless}). This property is closely related to the UV finiteness of $\mathcal{N}=2$ SCYM. Each individual one-loop determinant in the background field $\left\langle \Phi\right\rangle$ diverges in the UV. In (\ref{def:H}) this divergence is regularized in an arbitrary way, but in a more general setting, for instance in the $\mathcal{N}=2$ theory with $N_f\neq 2N$, the coefficient in front of $a^2$ in the effective action will be logarithmically divergent and will require adding a counterterm that shifts $1/\lambda$ and results in the one-loop beta-function. In $\mathcal{N}=2$ SCYM, any term quadratic in $a$ cancels due to the shift symmetry (\ref{shiftsymm}), which in particular means that the beta function is zero. It also means that there are no 
finite corrections to the coefficient in front  of $a^2$. The absence of finite renormalization has interesting consequences for the Wilson loop vev. The first renormalized vertex in the effective action  is thus quartic. The
lowest-order diagram with a quartic vertex  contains three propagators (fig.~\ref{fig:diagrams}) and thus contributes to the Wilson loop vev at $O(\lambda^3)$. The lower orders are described by the  Gaussian matrix model and thus are the same as in the $\mathcal{N}=4$ super-Yang-Mills, in agreement with the analysis of the SCYM perturbation theory \cite{Andree:2010na}.

The circular Wilson loop in SCYM on $S^4$ is a chiral observable and therefore
is directly related to Wilson loops (matrix exponentials) in the matrix model (\ref{Pestunmatrix}):
\begin{equation}\label{Wilson}
  W\left(C_{\rm circle}\right)=\left\langle \frac{1}{N}\sum_{i}^{}\,{\rm e}\,^{2\pi a_i}\right\rangle.
\end{equation}

\subsection{Saddle-point equation}

In the large-$N$ limit the saddle-point approximation becomes exact for the integral (\ref{Pestunmatrix}). The effective action for the eigenvalues is
\begin{equation}
 S(a)=\sum_{i}^{}\left(\frac{8\pi ^2}{\lambda }\,a_i^2+2\ln H(a_i)\right)
 -\frac{1}{N}\sum_{i<j}^{}\left(\ln\left(a_i-a_j\right)^2+2\ln H(a_i-a_j)\right).
\end{equation}
Minimizing the action we get the saddle-point equations:
\begin{equation}\label{spoint}
 \frac{8\pi ^2}{\lambda }\,a_i-K(a_i)-\frac{1}{N}\sum_{j\neq i}^{}
 \left(\frac{1}{a_i-a_j}-K\left(a_i-a_j\right)\right)=0.
\end{equation}
The function $K(x)$ that appears here is defined as 
\begin{equation}
 K(x)=-\frac{H'(x)}{H(x)}
 \end{equation}
 and is given  by
\begin{equation}\label{K(z)}
 K(x)=2x\sum_{n=1}^{\infty }\left(
 \frac{1}{n}-\frac{n}{n^2+x^2}
 \right)=x\left(\psi \left(1+ix\right)+\psi \left(1-ix\right)-2\psi (1)\right),
\end{equation}
where $\psi (x)$ is the logarithmic derivative of the Gamma-function: $\psi (x)=\Gamma '(x)/\Gamma (x)$.
The symmetry (\ref{shiftsymm}) translates into invariance of the saddle-point equations under the shifts of $K(x)$ by a linear function. We have used this symmetry to normalize $K(x)$ such that its Taylor expansion starts with $O(x^3)$.

As usual, the saddle-point equation can be interpreted as an equilibrium condition for $N$ pairwise interacting particles in the common external potential. Since
\begin{equation}\label{Kweak1}
K(x)\approx 2\zeta(3)x^3\qquad (x\rightarrow 0),
\end{equation}
the $K(x)$ terms in the force are negligible at short distances, and the equilibrium distribution is determined by the balance of the attractive harmonic potential and the $1/x$ pairwise repulsion. At large distances, on the contrary, $K(x)$ is large:
\begin{equation}
 K(x)\approx 2x\ln x \qquad \left(x\rightarrow +\infty \right),
\end{equation}
so the total one-body potential becomes repulsive, while the two-body interaction becomes attractive, exactly opposite to the short-distance behavior. The balance with this configuration of forces is still possible, but   the system becomes potentially unstable towards the spread of the eigenvalues to infinity. The stability again follows from mutual cancellation between the one-body $K(a_i)$  and two-body $K(a_i-a_j)$ terms for each individual particle, such that at large distances the harmonic potential still plays the most important role and confines the eigenvalue distribution to a finite interval.
The cancelations  have basically the same origin as the UV finiteness of theory.

We can introduce the eigenvalue density,
\begin{equation}
 \rho (x)=\frac{1}{N}\sum_{i}^{}\delta \left(x-a_i\right),
\end{equation}
which is defined on some interval $(-\mu ,\mu )$ and is unit normalized.
The saddle point equations take the form of a singular integral equation:
\begin{equation}\label{integralequation}
 \strokedint_{-\mu }^{\mu } dy\,\rho (y)\left(\frac{1}{x-y}-K(x-y)\right)=\frac{8\pi ^2}{\lambda }\,x-K(x),
\end{equation}
which together with the normalization condition determines the density and the endpoint $ \mu$.
The Wilson loop expectation value is given by the Laplace transform of the density:
\begin{equation}\label{WLoop}
 W(C_{\rm circle})=\int_{-\mu }^{\mu }dx\,\rho (x)\,{\rm e}\,^{2\pi x}.
\end{equation}

We  will develop systematic expansions of the saddle-point equation (\ref{integralequation}) at weak and at strong coupling. To this end, it will prove useful to rewrite the equation in a different form, suggested by the solution of the  Hermitean one-matrix model \cite{Brezin:1977sv} (which corresponds to setting $K(x)$ to zero). In this case the integral operator in (\ref{integralequation}) can be inverted by application of
$$
 \strokedint_{-\mu }^{\mu }\frac{dx}{\sqrt{\mu ^2-x^2}}\,\,\frac{1}{z-x}
$$
to both sides of the equation. If we apply this operator to (\ref{integralequation}), we get:
\begin{equation}\label{Fredholm}
 \rho (x)=\frac{8\pi }{\lambda }\,\sqrt{\mu ^2-x^2}-
 \frac{1}{\pi ^2}\strokedint_{-\mu }^{\mu }\frac{dy}{x-y}\,\,\sqrt{\frac{\mu ^2-x^2}{\mu ^2-y^2}}\int_{}^{}dz\,\rho (z)\left(K(y-z)-K(y)\right).
\end{equation}
We thus reduce the problem to an integral equation of the Fredholm type, since the kernel is not singular any more. The endpoints of the eigenvalue distribution are determined by the normalization condition:
\begin{equation}
 1=\frac{4\pi ^2\mu ^2}{\lambda }+\frac{1}{\pi }\int_{-\mu }^{\mu }
 \frac{dy\,y}{\sqrt{\mu ^2-y^2}}\,\,
 \int_{}^{}dz\,\rho (z)\left(K(y-z)-K(y)\right).
\end{equation}

\section{Weak coupling}\label{sec:weak}

In the weak coupling regime when $\lambda\ll 1$,  it follows immediately from the saddle point equation (\ref{integralequation}) that the eigenvalues are distributed on an interval $(-\mu ,\mu )$ with  $\mu\ll 1$.  In this regime, we can therefore express the function  $K(x)$  by the Taylor series
\begin{equation}\label{Kseries}
 K(x)=-2\sum_{n=1}^{\infty}(-1)^n\zeta(2n+1)x^{2n+1}
\end{equation}
where $\zeta(n)$ is the Riemann zeta function. At  the lowest order  of approximation we truncate the series by keeping only the first term (\ref{Kweak1}).

\begin{figure}[t]
\centerline{\includegraphics[width=10cm]{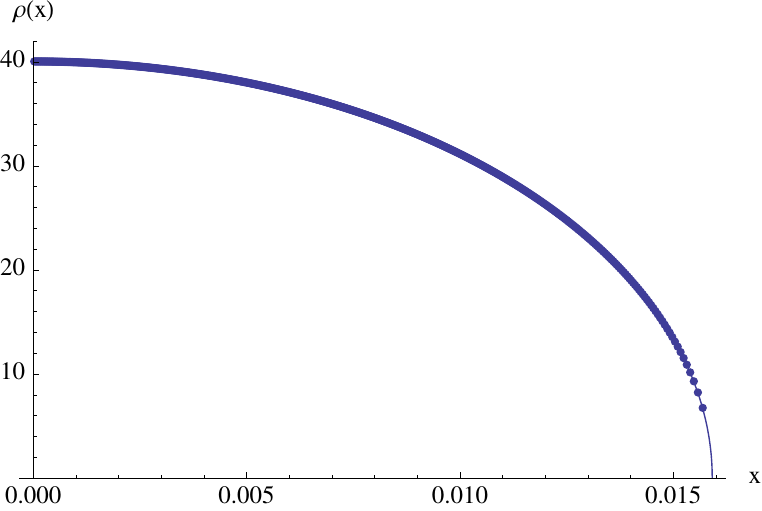}}
\caption{\label{pic:rho001}\small The weak coupling density at   $\lambda=0.01$ (solid line), compared to the numerical solution of the saddle-point equation with $N=1200$, shown in circles. }
\end{figure}

The last term in the equation  (\ref{Fredholm}) then factorizes, the integrals over $z$ and $y$ can be done separately, and
give  an  approximated expression for the density
\begin{equation}\label{FredholmWeak1}
 \rho (x)=\left(\frac{8\pi }{\lambda }\,+
\frac{6 \zeta(3)m_2}{\pi}\right)\sqrt{\mu ^2-x^2}\, ,
\end{equation}
where  we  defined  the second  moment
\begin{equation}
m_2=\int_{-\mu}^{\mu}dz\, \rho (z) z^2=\langle z^2 \rangle\, .
\end{equation}
The moment $m_2$ is self-consistently determined by the  density (\ref{FredholmWeak1}), and we obtain the equation
\begin{equation}\label{m2}
m_2=\left(\frac{8\pi }{\lambda }\,+
\frac{6 \zeta(3)m_2}{\pi}\right)\frac{\pi \mu^4 }{8}
\end{equation}
that permits us to express $m_2$   in terms of $\mu$ and $\lambda$.   Therefore,  at this order of  approximation, the density and the normalization condition are given explicitly  by
\begin{eqnarray}\label{RhoWeak1}
 \rho (x)&=&\left(\frac{8\pi }{\lambda }\,+
\frac{24 \pi  \zeta (3) \mu ^4}{4
   \lambda -3 \zeta (3) \lambda  \mu
   ^4}\right)\sqrt{\mu ^2-x^2}\\
   1&=&\frac{4\pi ^2\mu ^2}{\lambda }+\frac{12 \pi^2  \zeta (3) \mu ^6}{4
   \lambda -3 \zeta (3) \lambda  \mu
   ^4}
\end{eqnarray}
and  the normalization condition can be used to express $\mu$ as a function of $\lambda$. It results in 
\begin{equation}\label{MuLambda1}
\mu=\frac{\sqrt{\lambda }}{2 \pi }-\frac{3 \zeta (3) \lambda ^{5/2}}{256 \pi ^5}+\ldots
\end{equation}
The weak coupling  solution (\ref{RhoWeak1})  is compared to the numerical data in  fig.~\ref{pic:rho001}.
Using the density (\ref{RhoWeak1})  and the expression (\ref{MuLambda1}) it is possible to compute the expectation value of the circular Wilson loop,  as described in  (\ref{WLoop}).  We obtain
\begin{equation}\label{WLWeak1}
 W(C_{\rm circle})=1+\frac{\lambda }{8}+\frac{\lambda ^2}{192}+\left(\frac{1}{9216}-\frac{3 \zeta (3)}{512 \pi ^4}\right) \lambda^3+\ldots
\end{equation}
that  is in agreement with the result  of \cite{Andree:2010na}, considering the large $N$ limit.

The perturbative scheme just  outlined can be pushed to an arbitrary high order in $\lambda$. In particular, considering an expansion for  $K(z)$  up to order $O(z^{2M+1})$, the Fredholm equation (\ref{Fredholm}) gives an approximated  expression for the density that depends on $M$ moments $m_{2i}$ with $i=1,\ldots, M$. This density can be used to compute the moments, giving a system of $M$ equations that is the higher order generalization of the  (\ref{m2}) and that permits us to express the moments as  functions of $\mu$ and $\lambda$. In this way the density and the Wilson loop can be computed to an arbitrary order in $\lambda$. Expanding $K(z)$  up to order $O(z^{2M+1})$,  it  is possible to compute the expectation value of the Wilson loop up to order $O(\lambda^{2+M})$. In the appendix \ref{HOweak} this scheme is pushed to the seven-loop order for the Wilson loop vev.

\section{Strong coupling}\label{sec:strong}

The strong-coupling behavior of $\mathcal{N}=2$ SCYM on $S^4$ was analyzed in \cite{Rey:2010ry}. Using simple scaling arguments the authors of \cite{Rey:2010ry} reached the conclusion that the eigenvalue density approaches a finite limiting shape at $\lambda \rightarrow \infty $. Let us denote this limiting shape by $\rho _\infty (x)$. This behavior is in sharp contradistinction to the Gaussian matrix model which describes $\mathcal{N}=4$ super-Yang-Mills \cite{Erickson:2000af,Drukker:2000rr}, where the density is a function of the scaling variable $x/\sqrt{\lambda }$. For the $\mathcal{N}=2$ SCYM matrix model no consistent scaling is possible, as shown in \cite{Rey:2010ry}, and therefore the density freezes out as $\lambda \rightarrow \infty $. On the other hand, the weight in the  integral representation of the Wilson loop (\ref{WLoop}) exponentially grows with $x$, and thus the Wilson loop vev is determined  by the largest possible eigenvalue, $x=\mu $: $W\sim \,{\rm e}\,^{2\pi \mu }$ \cite{Rey:2010ry}. In the $\mathcal{N}=4$ context this leads to an exponential growth of the Wilson loop vev with $\sqrt{\lambda }$, and confirms the square-root dependence of the string tension on the 't~Hooft coupling (\ref{N=4}). The edge behavior of the density determines the prefactor in (\ref{WT}): the factor of $T^{-3/2}$ arises as a consequence of  the square-root singularity at the endpoint of the eigenvalue distribution \cite{Rey:2010ry}.
If in $\mathcal{N}=2$ SCYM the endpoint approaches a finite limiting value $\mu _\infty $, the Wilson loop vev also approaches a constant, which is really difficult to reconcile with the holographic duality. 

However, assuming that $\mu _\infty $ is a finite constant that does not depend on $\lambda $ we immediately run into contradiction. The integral equation (\ref{integralequation}) holds on the whole interval $(-\mu ,\mu )$ and in particular should be satisfied at  $x=\mu $. Assuming that the density freezes out at $\lambda \rightarrow \infty $ we can set $\lambda =\infty $ in the equation and then at the endpoint we get
$$
 \int_{-\mu _\infty }^{\mu _\infty }dy\,\rho _\infty (y)\left(
 K(\mu _\infty )-K(\mu _\infty -y)+\frac{1}{\mu_\infty -y }
 \right)=0,
$$ 
where we have used the normalization condition to move the first term inside the integral. The integral here converges without any regularization, because the density goes to zero at the enpoint as a square root of the distance: $\rho _\infty (y)\sim  \sqrt{\mu _\infty -y}$. But $K(x)$ is a monotonically growing function, and hence the integrand is strictly positive, so the equality can never be satisfied.

What can resolve this paradox? We must assume that the density approaches a limiting shape at $\lambda =\infty $, otherwise we run into contradiction with the scaling arguments of \cite{Rey:2010ry}. The only way to reconcile the existence of the limiting density with the saddle-point equation is to admit that $\mu _\infty =\infty $. This behavior is extremely unusual for matrix models. It is ultimately related to the intrinsic instability of balancing the one-body repulsion against the two-body attraction, as we  discussed in sec.~\ref{sec:partition}. We have performed extensive numerical checks of this behavior by solving the saddle-point equations (\ref{spoint})  at large but finite $N$ and $1/\lambda =0$. In the numerics we have dealt with finitely many eigenvalues  which of course do not extend to infinity,
but the largest eigenvalue, as we have found, grows  more or less linearly with $N$ without any signs of saturation. 

\subsection{Infinite coupling}

\begin{figure}[t]
\centerline{\includegraphics[width=10cm]{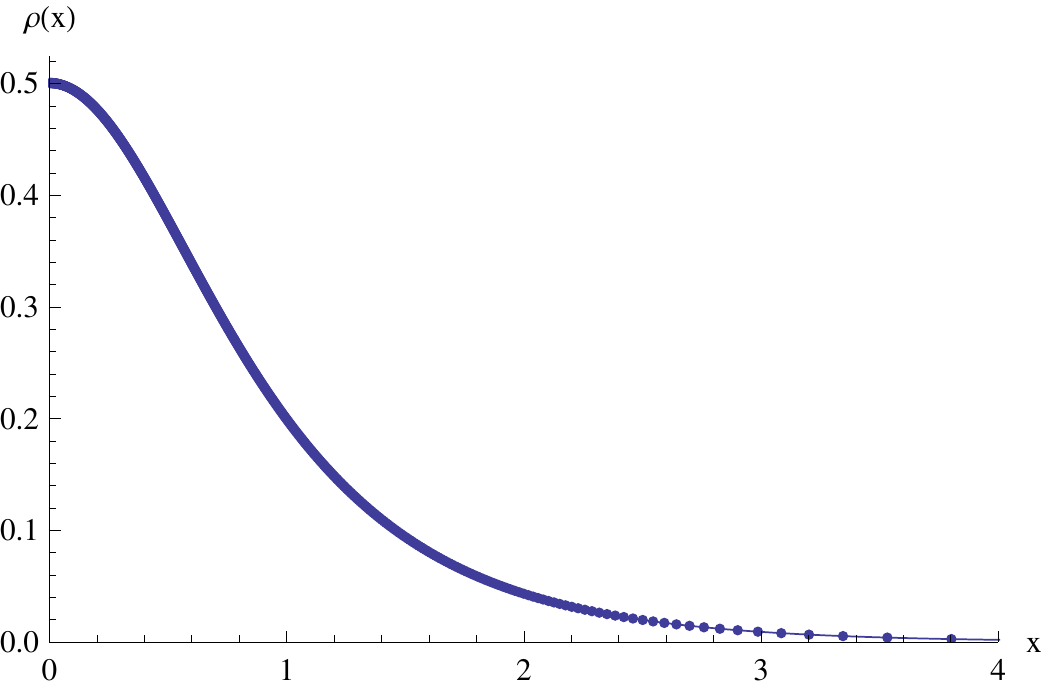}}
\caption{\label{pic:rhoinf}\small The density at infinite coupling (solid line), compared to the numerical solution of the saddle-point equation with $N=1200$ and $1/\lambda =0$, shown in circles. }
\end{figure}

The equation for the limiting shape of the eigenvalue distribution is
\begin{equation}
 \strokedintmore_{-\infty  }^{+\infty  } dy\,\rho_\infty  (y)\left(\frac{1}{x-y}-K(x-y)\right)=- K(x).
\end{equation}
It can be solved by Fourier transform:
\begin{equation}\label{Fourier}
 \pi i\mathop{\mathrm{sign}}\omega \left[1+\frac{1}{2\sinh^2\frac{\omega }{2}}\right]\rho _\infty (\omega )
 =\frac{\pi i\mathop{\mathrm{sign}}\omega }{2\sinh^2\frac{\omega }{2}}\,,
\end{equation}
which gives:
\begin{equation}
 \rho _\infty (\omega )=\frac{1}{\cosh \omega }\,,
\end{equation}
and
\begin{equation}\label{Rhoinf}
 \rho _\infty (x)=\frac{1}{2\cosh\frac{\pi x}{2}}\,.
\end{equation}
In fig.~\ref{pic:rhoinf} this solution is compared to the numerical data.

The majority of eigenvalues are indeed concentrated at $x=O(1)$, as the scaling arguments suggest, but the density has exponential tails extending all the way to infinity. The shape of the density is such that the scaling arguments apply to an observable $\left\langle \mathcal{O}(x)\right\rangle$ if and only if $\mathcal{O}(x)$ does not grow with $x$ faster than $\,{\rm e}\,^{\pi x/2}$. The expectation value  $\left\langle \mathcal{O}(x)\right\rangle$ then approaches a constant limiting value at $\lambda \rightarrow \infty $. The Wilson loop does not belong to this class of observables, since the integral (\ref{WLoop}) with the limit density diverges. Therefore, the strong-coupling behavior of the Wilson loop is still determined by the largest eigenvalue and is sensitive to the precise value of $\mu $ and to the endpoint behavior of the eigenvalue density at large but finite $\lambda $. In order to compute the Wilson loop we need to analyze  deviations from the strict infinite-coupling limit. Technically this is a much more difficult problem, and we start with simple qualitative estimates before proceeding to a more systematic analysis.

\subsection{Qualitative estimates}

If $\lambda $ is large but finite, the density cannot differ much from $\rho _\infty (x)$. The most important difference is the finite extent of the eigenvalue distribution. The exponential tails of $\rho _\infty (x)$ are trimmed at $x=\pm\mu $ in a certain way, which we are going to analyze. As we have seen,  there is an intrinsic instability in the balance of forces at strong coupling. Because of this instability the effective force acting on an individual eigenvalue goes to zero at infinity (when $\lambda$ is strictly infinite), and this allows
the eigenvalues 
 to spread all over the real line. If $\lambda $ is not infinite  but very large, the linear term in the force can be neglected in a certain range of $x$, since it has a very small coefficient. But  eventually it takes over and starts to press the eigenvalues towards the origin.  As a result, the density becomes slightly bigger everywhere,  and cannot extend further than $x=\pm \mu $. As a crude approximation we can neglect the linear force at $x\sim 1$ and neglect the two-body forces at $x\sim \mu $. The eigenvalues density is then approximated by (\ref{Rhoinf}) at $x\sim 1$ and by the Wigner semi-circle at $x\sim \mu $:
\begin{equation}\label{Wig}
 \rho (x)\sim \frac{8\pi }{\lambda }\,\sqrt{\mu ^2-x^2}\,.
\end{equation}
Matching the two at some intermediate scale we find that the endpoint $\mu $ should scale logarithmically with $\lambda $:
\begin{equation}
  \mu = \frac{2}{\pi }\,\ln\lambda +\ldots \qquad (\lambda \rightarrow \infty ) .
\end{equation}

We can obtain a more accurate estimate by taking into account the normalization condition. The linear force eliminates all the eigenvalues that sit at $x>\mu $, compressing them to smaller $x$. Let us consider the positive endpoint $x=+\mu $.
The fraction of eigenvalues redistributed to smaller $x$ roughly speaking is given by
$$
 \int_{\mu }^{\infty }dx\,\rho _\infty (x)\simeq \frac{2}{\pi }\,\,{\rm e}\,^{-{\pi \mu }/{2}}.
$$
If we approximate the eigenvalue distribution near the endpoint by the Wigner semi-circle, the excess number of eigenvalues, compared to the in\-fi\-ni\-te-range case, is obtained by integrating the Wigner distribution from some cutoff scale $\mu -z_0$ ($z_0\sim 1$) to $\mu $:
$$
 \int_{\mu -z_0}^{\mu }dx\,\,\frac{8\pi }{\lambda }\,\sqrt{\mu ^2-x^2}\simeq
 \frac{16\pi \sqrt{2}\,z_0^{3/2}}{3\lambda }\,\sqrt{\mu }\,.
$$
Equating this with the number of eigenvalues that came from infinity, we get an equation
\begin{equation}\label{mularge}
 C\sqrt{\mu }\,{\rm e}\,^{\pi \mu /2}=\lambda ,
\end{equation}
which determines $\mu $ as a function of $\lambda $.
We cannot find the constant of proportionality in this equation from the simple arguments above (the constant depends on the cutoff $z_0$ which we have put in by hand).
To honestly compute this constant we need more sophisticated methods.
Later we will estimate 
\begin{equation}\label{numerics}
 {C} \simeq 14.60,
\end{equation}
which  corresponds to taking $z_0=0.54$ in the previous simple-minded argument, from which we can get an idea how well the actual density near the endpoint is approximated by the Wigner semi-circle.

\begin{figure}[t]
\centerline{\includegraphics[width=10cm]{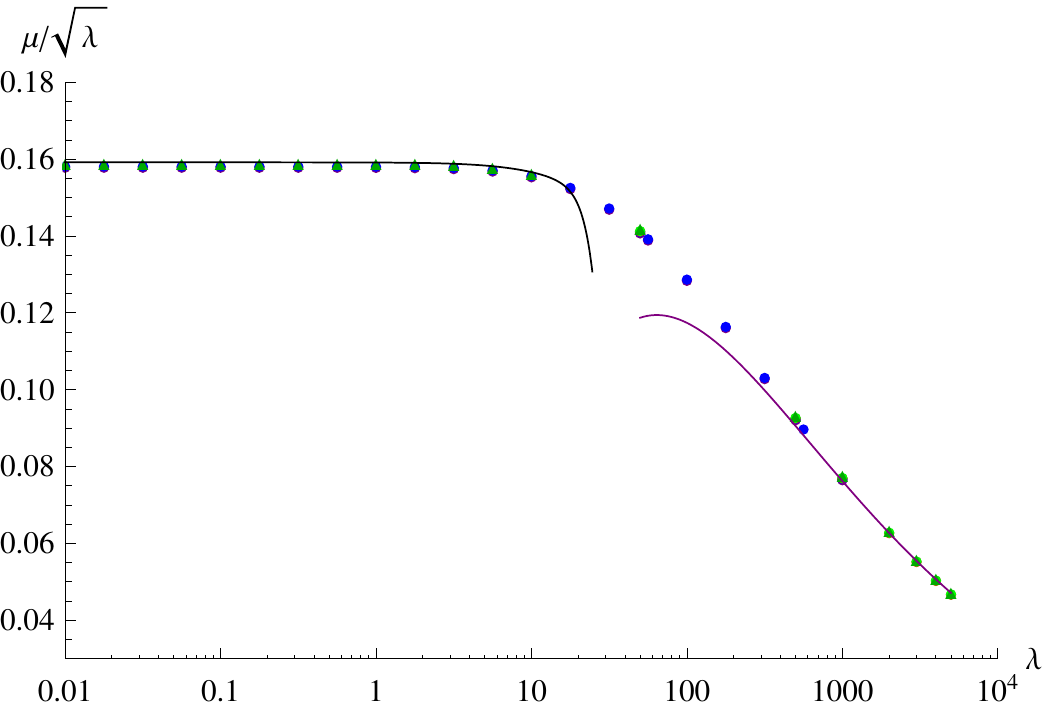}}
\caption{\label{pic:mulambda}\small The ratio $\mu/\sqrt{\lambda }$ as a function of $\lambda $: dots represent numerical results; black curve is the seven-loop weak-coupling approximation (\ref{ExpWeak3}); purple curve is the strong-coupling prediction (\ref{mularge}), (\ref{numerics}). }
\end{figure}

In fig.~\ref{pic:mulambda} we compare the weak-coupling and strong-coupling predictions for the endpoint position with the numerical data (see also appendix~\ref{app:endpoint}).

We can estimate the Wilson loop vev in a similar way\footnote{We would like to thank the anonymous referee of JHEP for suggesting this calculation to us.}. Here we may expect better accuracy, because the exponential weight of the Wilson loop is peaked near the positive endpoint of the eigenvalue distribution,  and this is precisely where it makes sense to use
the simple semi-circle approximation (\ref{Wig}):
\begin{equation}
 W(C_{\rm circle})\sim \frac{8\pi }{\lambda }\int_{}^{\mu }dx\,\sqrt{\mu ^2-x^2}\,\,{\rm e}\,^{2\pi x}.
\end{equation}
Doing the integral we find that the Wilson loop vev scales with $\lambda $ and $\mu $ as
\begin{equation}\label{vevscaling}
  W(C_{\rm circle})=R\,\frac{\sqrt{\mu }}{\lambda }\,\,{\rm e}\,^{2\pi \mu }
\end{equation}
with $R=2$. Later we will get a better estimate of the proportionality constant which, as a matter of fact, does not differ much from the simple-minded approximate calculation above:
\begin{equation}\label{Rfin}
 R=2.18.
\end{equation}

Expressing $\mu $ in terms of $\lambda $ with the help of (\ref{mularge}), we find from (\ref{vevscaling}) the results (\ref{llambdasca}), (\ref{WT}) quoted in the introduction. 
For the constant $K$ in (\ref{WT}) we get:
\begin{equation}
 K=\left(\frac{3}{4}\right)^{3/2}\,\frac{R}{C^4}\,,
\end{equation}
and using (\ref{numerics}), (\ref{Rfin}) we obtain the numerical estimate
(\ref{Kconst}).

The logarithmic dependence of the endpoint position on the 't~Hooft coupling may signal the logarithmic branch point at infinity, which then implies that other complex solutions may exist, related to the original real solution by a non-trivial monodromy at infinity. We have studied this question numerically by solving the saddle-point equations for complex $\lambda =|\lambda |\,{\rm e}\,^{i\phi }$ and slowly changing $\phi $ from $0$ to $2\pi $. We have found that the monodromy is actually trivial and that the saddle-point equation does not have complex solutions for real $\lambda $. When $\lambda $ acquires a non-zero phase, the eigenvalue distribution extends into the complex plane and, for small coupling, rotates in phase with $\sqrt{\lambda }$, such that for $\phi $ going from $0$ to $2\pi $ the eigenvalue cut makes a $180^\circ$ turn and maps to itself. The picture at strong coupling is more complicated but again the density maps to itself as soon as the phase of $\lambda $ changes by $2\pi $.

\subsection{Structure of the density}\label{structure}

When $\mu $ is large we can identify three regions in which the density has qualitatively different behavior. At $x\sim 1$, the density is well approximated by the infinite-coupling solution (\ref{Rhoinf}), and varies on distances of order one. We call this region I. At $x\sim \mu $, but not too close to the endpoints, the density changes very slowly. The scale of its variation is of order $\mu $. There we can introduce the scaling variable $x/\mu $. We call this region II. Finally, near the endpoints the density rapidly changes again on the scales of order one. We call this region III.

One can infer from (\ref{Fredholm}) that in the region III the density is proportional to $\sqrt{\mu }/\lambda $:
\begin{equation}\label{III}
 \rho (x)=\frac{\sqrt{\mu }}{\lambda }\,f(\mu-x )\qquad (\mu -x\sim 1),
\end{equation}
which is also consistent with the simple estimates in the previous section. The scaling function behaves as $f(z)\sim \sqrt{z}$ at $z\rightarrow 0^+$. As we shall see later, $f(z)\sim z^{-3/2}$ at $z\rightarrow \infty $. An integral equation for $f(z)$ of Fredholm type can be obtained by plugging (\ref{III}) in (\ref{Fredholm}) and taking $\mu \rightarrow \infty $ while keeping $\mu -x$ fixed. We will not study this equation here.

To solve for the density in region II it is convenient to use yet another integral representation for  $K(x)$:
\begin{equation}\label{anotherIR}
 K(x)=\strokedintmore_{-\infty  }^{+\infty  }
 \frac{dw\,w\coth\pi w}{x-w}\,.
\end{equation}
The integral diverges at large $w$ and thus this dispersion relation  only holds after two subtractions. Namely, to make it precise we first need to shift the argument of $K(x)$ by $z$ and average over $z$ such that the first two moments vanish. But this is precisely how $K(x)$ appears in the Fredholm representation (\ref{Fredholm}).

Upon substitution of (\ref{anotherIR}) into (\ref{Fredholm}) the double integral in $y$ and $w$ essentially acts as a unit operator as long as $w\in(-\mu ,\mu )$, because of the principal value prescription to encircle the poles. The remainder integrates to zero at $|w|<\mu $. After some transformations, we get
\begin{eqnarray}\label{ineq}
 &&\rho (x)-\int_{}^{}dy\,\rho (y)(x-y)\coth\pi (x-y)+x\coth\pi x
 =\frac{8\pi }{\lambda }\,\sqrt{\mu ^2-x^2}
\nonumber \\
&&
 -\frac{1}{\pi }\int_{|w|>\mu }^{}\frac{dw}{w-x}\,\,\sqrt{
 \frac{\mu ^2-x^2}{w^2-\mu ^2}}\int_{}^{}dy\,\rho (y)\left[
 \left(w-y\right)\coth\pi \left(w-y\right)-w\coth\pi w
 \right].\nonumber\\
\end{eqnarray}
The advantage of this representation is that the last line is always small.
For the major part of the eigenvalue distribution $|x|\lesssim \mu $. Hence $1/(w-x)$ in the integrand is suppressed by a factor of order $1/\mu $. In addition, for large $x$:
\begin{equation}\label{|x|}
 x\coth\pi x\approx |x|.
\end{equation}
Assuming, for definiteness, positive $w$ and taking into account that then $w-y$ is also positive, we get
\begin{eqnarray}
  &&\int_{}^{}dy\,\rho (y)\left[
 \left(w-y\right)\coth\pi \left(w-y\right)-w\coth\pi w
 \right]
\nonumber \\ 
\nonumber
 &&\approx \int_{}^{}dy\,\rho (y)\left(w-y-w
 \right)=0.
\end{eqnarray}
The last term in (\ref{ineq}) is thus localized in a small vicinity of the distribution's endpoints, where $\mu -x$, $w-\mu $ and $\mu -y$ are all small compared to $\mu $. We can safely ignore this term in regions I and II, and we will also argue that this term is numerically small in region III. Then the equation (\ref{ineq}) becomes
\begin{equation}\label{cotheq}
 \rho (x)-\int_{-\mu }^{\mu }dy\,\rho (y)(x-y)\coth\pi (x-y)+x\coth\pi x
 = \frac{8\pi }{\lambda }\,\sqrt{\mu ^2-x^2}\,.
\end{equation}

In region I, at $x\sim 1$, we can neglect the last term in (\ref{cotheq}) and solve the equation by the Fourier transform, which gives the asymptotic solution (\ref{Rhoinf}).  In region II, the last term can no longer be neglected, but the density slowly varies with $x$ and consequently the typical range of $y$'s that contributes to the integral on the left-hand side is large, such that $x-y$  is of order $\mu $. We can then use the long-range approximation for the kernel (\ref{|x|}). Differentiating the resulting equation twice and taking into account that $|x|''=2\delta (x)$, we find:
\begin{equation}
 \rho ''(x)-2\rho (x)+2\delta (x)=-\frac{8\pi \mu ^2}{\lambda }\,\left(\mu ^2-x^2\right)^{-3/2}.
\end{equation}
The delta-function can be dropped, as $x$ cannot be close to zero in region II. Each derivative brings in a factor of $1/\mu $, since  $x$ scales as $\mu $, and we thus have:
\begin{equation}\label{RII}
 \rho (x)=\frac{4\pi \mu ^2}{\lambda }\,\left(\mu ^2-x^2\right)^{-3/2}
\qquad (x\sim \mu ).
\end{equation}
Matching this to the solution (\ref{III}) in region III, we find that the function $f(z)$ defined there behaves at large $z$ as
\begin{equation}
 f(z)\approx \frac{\sqrt{2}\pi }{z^{3/2}}\qquad (z\rightarrow \infty ).
\end{equation}

The solution (\ref{RII}) is non-normalizable, because it has a wrong endpoint behavior: $(\mu -x)^{-3/2}$ instead of $(\mu -x)^{1/2}$. To compute the normalization integral and thus to determine $\mu $ as a function of $\lambda $ we need to compute the density in region III and also to match the solutions in regions I and II. For that we need to solve eq.~(\ref{cotheq}) for $\mu \gg 1$.  
For $\mu =\infty $ we could have used the Fourier transform. For large but finite interval, the solution can be obtained by a generalization of the Wiener-Hopf method \cite{Ganin,Yang:1966sa}. 

\subsection{Wiener-Hopf solution}

The Wiener-Hopf method is a generalization of the Fourier transform for the case  when an integral equation is defined on a finite interval. The idea of the method is to focus on the vicinity of one end-point, and make sure that the boundary conditions there are correct, at the same time neglecting the influence of the other endpoint.  This will give an accurate description of the density in all regions I--III  for $x>0$, as long as we impose the correct boundary conditions at $x=\mu $, while the solution will give a bad approximation to the density in the region III at $x<0$ (in the vicinity of $x=-\mu $). But since the exact density is an even function of $x$ it it sufficient to know it for positive $x$. In particular, this approximation will be sufficient for the computation of the Wilson loop (\ref{WLoop}), since it is dominated by $x$ close to $+\mu $.

The Wiener-Hopf method is based on the analytic decomposition of the kernel in the integral equation:
\begin{equation}\label{GtoG+G-}
  \frac{\cosh\omega }{2\sinh^2\frac{\omega }{2}}=\frac{1}{G_-(\omega )G_+(\omega )}\,,
\end{equation}
where the functions
\begin{equation}\label{kerns}
 G_\pm(\omega )=\frac{\sqrt{8\pi^3 }\,2^{\pm i\omega /\pi }\Gamma \left(\frac{1}{2}\mp\frac{i\omega }{\pi }\right)}{\omega \Gamma^2 \left(\mp\frac{i\omega }{2\pi }\right)}
\end{equation}
are analytic on the upper/lower half-plane.

\begin{figure}[t]
\centerline{\includegraphics[width=8cm]{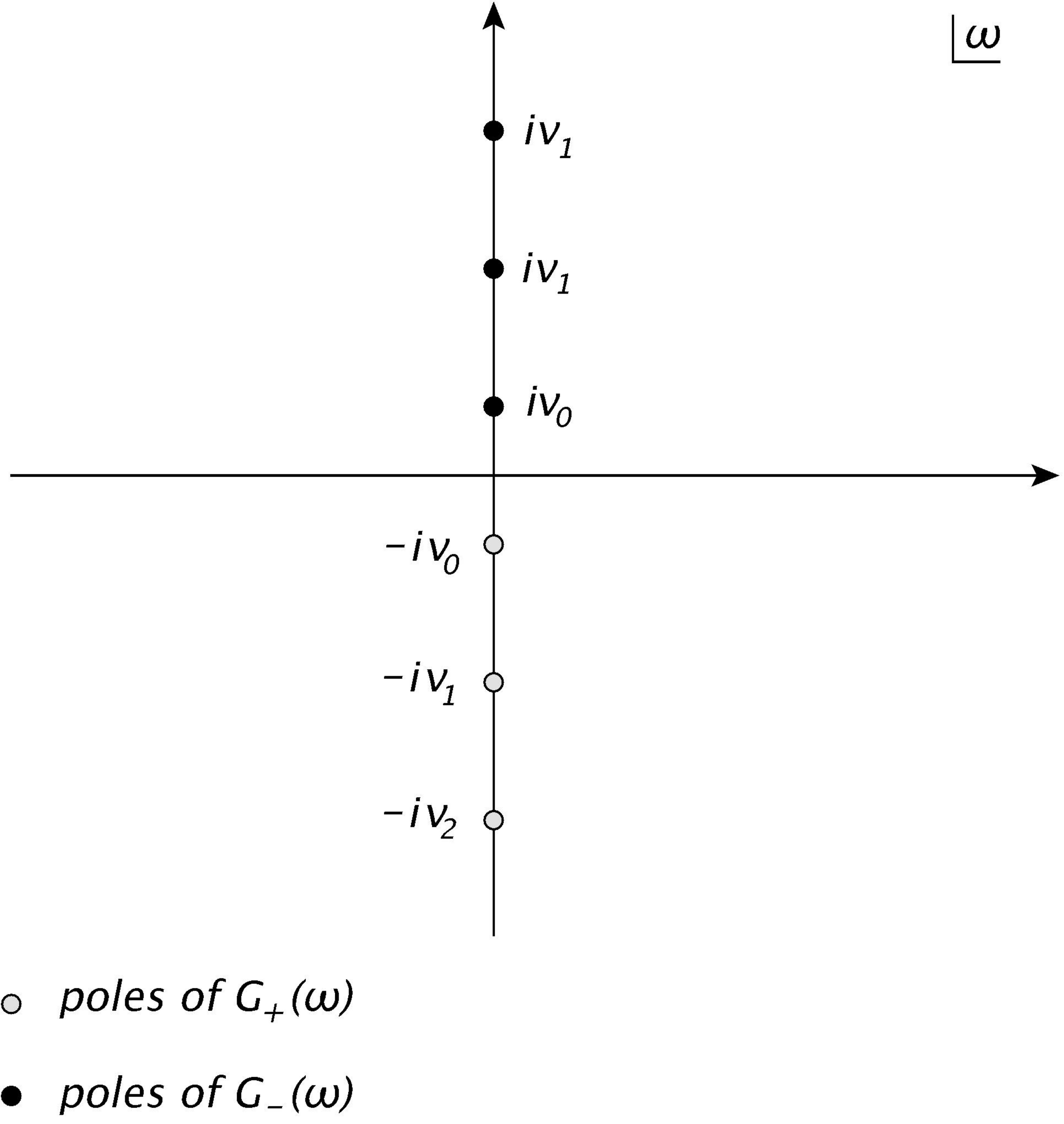}}
\caption{\label{astr}\small The analytic structure of $G_\pm(\omega )$. }
\end{figure}

The analytic properties of the Wiener-Hopf kernels are illustrated in fig.~\ref{astr}.
The only singularities of $G_\pm(\omega )$ are simple poles at $\omega =\mp i\nu _n$ with
\begin{equation}\label{poles}
 \nu _n=\pi \left(n+\frac{1}{2}\right).
\end{equation}
The residues are 
\begin{equation}\label{residues}
 r_n\equiv \mathop{\mathrm{res}}_{\omega =\mp i\nu _n}G_\pm(\omega )
 =\frac{\left(-2\right)^{n+1}\Gamma ^2\left(\frac{n}{2}+\frac{5}{4}\right)}{\sqrt{\pi }\left(n+\frac{1}{2}\right)\Gamma \left(n+1\right)}\,.
\end{equation}
The kernels satisfy
\begin{equation}
 G_\pm(\bar{\omega })=\bar{G}_\mp(\omega ).
\end{equation}

The Wiener-Hopf solution of the integral equation, which is accurate at positive $x$, in the Fourier space is given by
\begin{equation}\label{densityfin}
 \rho (\omega )=\frac{1}{\cosh\omega }+\frac{2\sinh^2\frac{\omega }{2}}{\cosh\omega }\,F(\omega )+G_-(\omega )\,{\rm e}\,^{i\mu \omega }\sum_{n=0}^{\infty }\frac{r_n\,{\rm e}\,^{-\mu \nu _n}}{\omega +i\nu _n}\left(1-F(-i\nu _n)\right),
\end{equation}
where $F(\omega )$ is the Fourier transform of the right-hand-side of the integral equation (\ref{ineq}).
The technical details of the derivation are collected in the appendix~\ref{app:wienerhopf}. 

The first term in (\ref{densityfin}) is the asymptotic solution on the infinite interval. The second term is a correction due to the linear force, which in particular gives (\ref{RII}) in the region II. To see this, we can replace $2\sinh^2(\omega /2)/\cosh\omega$ by $\omega ^2/2\rightarrow -\partial ^2/2$, which is justfied because the density varies very slowly in region II. The density then is given by the second derivative of the Wigner's semi-circle, in accord with (\ref{RII}). 

\begin{figure}[t]
\centerline{\includegraphics[width=10cm]{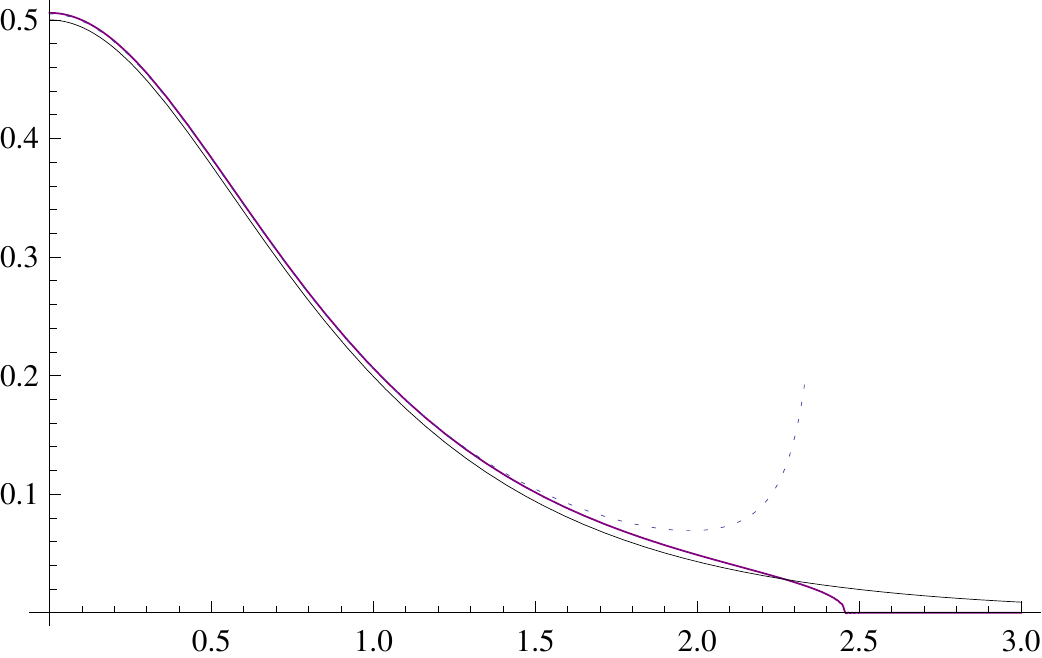}}
\caption{\label{pic:density}\small The $x$-space density $\rho(x)$ at $\lambda=1000$ and $\mu=2.4546$ (solid purple line), compared to the infinite coupling solution $\rho_\infty(x)$ (solid black line) and  to  $\rho_\infty(x)+\frac{4\pi \mu ^2}{\lambda }\,\left(\mu ^2-x^2\right)^{-3/2}$, that is the region II density (dashed black line).}
\end{figure}

\begin{figure}[t]
\centerline{\includegraphics[width=10cm]{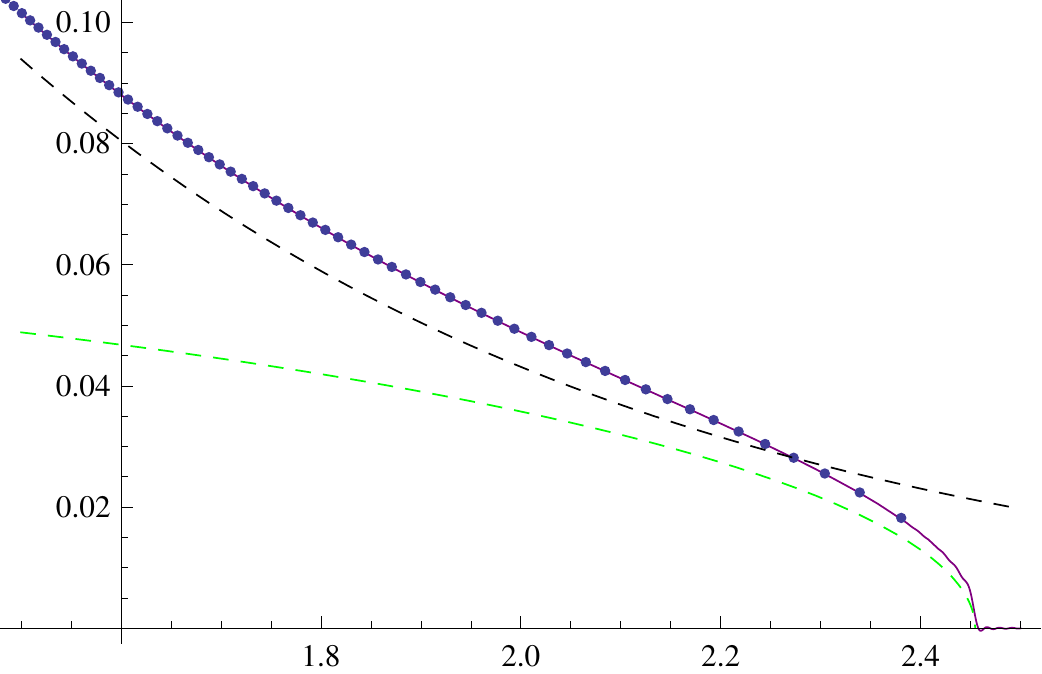}}
\caption{\label{pic:edge}\small The $x$-space density $\rho(x)$ at $\lambda=1000$ and $\mu=2.4546$ near  $x=\mu$  (solid purple line), compared to the infinite coupling solution $\rho_\infty(x)$ (dashed black line), to   the Wigner semi-circle (dashed green line) and to the numerical solution of the saddle-point equation with $N=1200$ (circles).}
\end{figure}

The last term in (\ref{densityfin}) subtracts the the poles of the first two terms at $\omega =-i\nu _n$ thus ensuring that
 $\rho (\omega )$ is analytic in the lower half-plane. This means that the $x$-space density vanishes at $x>\mu $ and the solution thus satisfies the correct boundary conditions at $x=\mu $. In fig.~\ref{pic:density}  we have plotted  the $x$-space density $\rho(x)$. For comparison we also plot the infinite coupling distribution (\ref{Rhoinf}) and the approximate solution in the region II (\ref{RII}). In  fig.~\ref{pic:edge}  the $x$-space density near  $x=\mu$  is compared to the numerical data, to the infinite coupling solution and to the Wigner semi-circle (\ref{Wig}).

To find $\mu $ as a function of $\lambda $ we need to impose the normalization condition on the density. We cannot do it by requiring that $\rho (\omega =0)=1$, because the solution (\ref{densityfin}) is approximate and gives a good approximation to the density only at $x>0$. In particular,  $\rho (x<-\mu )\neq 0$. $\rho (\omega =0)$ is given by the integral of the approximate solution from $-\infty $ to $+\infty $ (effectively to $\mu $, since we imposed the right boundary conditions at $x=\mu $ and for $x>\mu $ the approximate density vanishes). Because the measure of integration is flat, the integral picks a contribution from the region of negative $x$ where the solution  does not approximate the true density with sufficient precision. 

We should instead calculate the normalization as
\begin{equation}
 1=2\int_{0}^{\mu }dx\,\rho (x)=\int_{-\infty }^{+\infty }\frac{d\omega }{\pi i}\,\,\frac{\rho (\omega )}{\omega -i0}\,,
\end{equation}
using the symmetry of the exact solution under $x\rightarrow -x$. This trick allows us to avoid using the approximate solution beyond the region where it gives a controllable approximation.
Plugging in the explicit expression (\ref{densityfin}) we find:
\begin{equation}
 1=1+2\sum_{n,m=0}^{\infty }\frac{r_mr_n\,{\rm e}\,^{-\mu \left(\nu _m+\nu _n\right)}}{\nu _m\left(\nu _m+\nu _n\right)}\,\left(F(-i\nu _n)-1\right)
\end{equation}
The sum over $m$ here is dominated by the first term. The rest are  exponentially small in  $ \mu$. However, $F(-i\nu _n)$ is exponentially large, $O(\,{\rm e}\,^{\mu \nu _n})$, and thus all the terms in $n$ should be kept. This gives the normalization condition:
\begin{equation}\label{sums}
 \sum_{n=0}^{\infty }\frac{r_n\,{\rm e}\,^{-\mu \nu _n}F(-i\nu _n)}{\nu _n+\nu _0}=\frac{r_0}{2\nu _0}\,\,{\rm e}\,^{-\mu \nu _0}.
\end{equation}
With the help of this condition, the density (\ref{densityfin}) can be rewritten in the form that sometimes will be more convenient:
\begin{eqnarray}\label{densityfin1}
 \rho (\omega )&=&\frac{1}{\cosh\omega }+\frac{2\sinh^2\frac{\omega }{2}}{\cosh\omega }\,F(\omega )
\nonumber \\
&&-G_-(\omega )\,{\rm e}\,^{i\mu \omega }\sum_{n=1}^{\infty }\frac{r_n\,{\rm e}\,^{-\mu \nu _n}F(-i\nu _n)\left(\nu _n-\nu _0\right)\left(\omega -i\nu _0\right)}{\left(\nu _n+\nu _0\right)\left(\omega +i\nu _0\right)\left(\omega +i\nu _n\right)}.
\end{eqnarray}

As a first approximation we can neglect the second line in (\ref{ineq}) and start with (\ref{cotheq}). Then
\begin{equation}
 F(x)=\frac{8\pi }{\lambda }\,\sqrt{\mu ^2-x^2}\,,
\end{equation}
and
\begin{equation}\label{Fomega}
 F(\omega )=\frac{8\pi ^2\mu J_1(\mu \omega )}{\lambda \omega }\,,
\end{equation}
where $J_1(x)$ is the Bessel function of the first kind. In the lower half-plane we get:
\begin{equation}\label{largemu}
 F(-i\nu _n)=\frac{8\pi ^2\mu I_1\left(\mu \nu _n\right)}{\lambda \nu _n}
 \approx
 \frac{\,{\rm e}\,^{\mu \nu _n}}{\lambda }\,
 \sqrt{\frac{32\pi ^3\mu }{\nu _n^3}}\,.
\end{equation}
Substituting this into eq.~(\ref{sums}) we find that the normalization condition results in the equation (\ref{mularge}) with  the constant
$C$ given by
\begin{equation}\label{lo}
 C=\frac{16}{r_0}\sum_{n=0}^{\infty }\frac{r_n}{\left(2n+1\right)^{3/2}\left(n+1\right)}=15.16.
\end{equation}

The first correction due to the second line in (\ref{ineq}) is computed in the appendix~\ref{1interation}. Parametrically it is of the same order, but numerically turns out to be rather small:  the correction to the normalization constant is $\delta C=-0.56$. The estimate (\ref{numerics}) is a combination of the leading-order result (\ref{lo}) and this correction. We expect that higher-order corrections are at least as small as the first one, which means that we know the constant $C$ at a few percent accuracy. This is consistent with numerics, as discussed in appendix~\ref{app:endpoint}.

\subsection{Wilson loop}

The Wilson loop can be computed as\footnote{The right-hand side is the Fourier transform of the density evaluated at $\omega =-2\pi i$.}
\begin{equation}
 W(C_{\rm circle})=\rho (-2\pi i).
\end{equation}
When we plug in the solution (\ref{densityfin}), we immediately see that the exponentially growing term in the Wilson loop vev comes from the last term in (\ref{densityfin}) which describes the solution in region III. The Wilson loop is thus determined by the behavior of the density near the endpoint of the eigenvalue distribution, as expected on general grounds. Using (\ref{densityfin1}) for $\rho (\omega )$ and (\ref{kerns}) for $G_-(\omega )$, we get, omitting exponentially small terms:
\begin{equation}\label{Wilint}
 W(C_{\rm circle})=\,{\rm e}\,^{2\pi \mu }\,
\frac{5}{\sqrt{32} }
  \sum_{n=1}^{\infty }\frac{nr_n\,{\rm e}\,^{-\mu \nu _n}F(-i\nu _n)}{\left(n+1\right)\left(3-2n\right)}\,.
\end{equation}
It is clear from (\ref{largemu}) that the Wilson loop vev scales as (\ref{vevscaling}) with
\begin{equation}
 R={5}\sum_{n=1}^{\infty }\frac{nr_n}{\left(n+\frac{1}{2}\right)^{3/2}\left(n+1\right)\left(3-2n\right)}=2.55\,.
\end{equation}
The next-order correction to $R$  is computed in appendix~\ref{1interation} and is not very big: $\delta R=-0.37$. Altogether we get an estimate quoted in eq.~(\ref{Rfin}).

\section{Instantons}\label{sec:instantons}

Since instantons can lead to a large-$N$ phase transition
\cite{Gross:1994mr}, we will compute the large-$N$ limit of the one-instanton contribution to the partition function, in order to check if the moduli integration can overcome the exponential suppression of the instanton weight. 
 The total one-instanton weight is given by
\begin{eqnarray}
 \mathcal{Z}_{\rm 1-inst}&=&\,{\rm e}\,^{-\frac{8\pi ^2}{g^2}}\,
 2\mathop{\mathrm{Re}}
 \sum_{k=1}^{N}
 \frac{(a_k+i)^{2N}}{\prod\limits_{j\neq k}^{}\left(a_k-a_j\right)\left(a_k-a_j+2i \right)}
\nonumber \\
& =&4\,{\rm e}\,^{-\frac{8\pi ^2}{g^2}}\,
 \int_{-\infty }^{+\infty }\frac{dy}{2\pi}\,\left\{
\frac{y^{2N}}{\prod\limits_{j}
 \left[\left(y-a_j\right)^2+1\right]
 }-1\right\}.
\end{eqnarray}
The deformation parameters, that enter all instanton sums \cite{Nekrasov:2002qd,Nekrasov:2003rj}, according to \cite{Pestun:2007rz} should be set to one for the $\mathcal{N}=2$ partition function on $S^4$: $\varepsilon _1=\varepsilon _2=1$.
We have taken into account that the mass of the hypermultiplet is offset by $(\varepsilon _1+\varepsilon _2)/2=1$ \cite{Okuda:2010ke}, resulting in the factor of $(a_k+i)^{2N}$ in the numerator, which is the canonical contribution of $2N$ fundamental hypermultiplets of mass $1$ to the one-instanton weight.

In the large-$N$ limit the first term in the curly brackets is exponentially small as long as $y$ is not too big. Only for  $y$ of order $\sqrt{N}$ this term becomes sizable and we need to take it into account. We can thus approximate the integrand by expnading the exponent in the first term in $1/y$:
\begin{eqnarray}
 \frac{y^{2N}}{\prod\limits_{j}
 \left[\left(y-a_j\right)^2+1\right]
 }
 &\approx& 
 \exp\left\{{\sum_{j}^{}\left[\frac{2a_j}{y}-\frac{1-a_j^2}{y^2}+O\left(\frac{1}{y^3}\right)\right]}\right\}
\nonumber \\
\nonumber 
&=&\exp\left[{-N\,\frac{1-\left\langle a^2\right\rangle}{y^2}+O\left(\frac{1}{y^4}\right)}\right].
\end{eqnarray}
The $y$ integral becomes elementary in this approximation and we find for the one-instanton contribution, to the leading order in $1/N$:
\begin{equation}
 \mathcal{Z}_{\rm 1-inst}=-4\sqrt{\frac{1-\left\langle a^2\right\rangle}{\pi }\,N}\,{\rm e}\,^{-\frac{8\pi ^2N}{\lambda }}\,.
\end{equation}
We thus conclude that the moduli integration enhances the instanton weight by a factor of $\sqrt{N}$, but does not overcome the exponential suppression of the weight by the instanton action.
 
\section{Conclusions}\label{sec:conclusions}

At weak coupling our results are in agreement with the perturbative calculation of the Wilson loop from \cite{Andree:2010na}. An observation that the difference between the $\mathcal{N}=4$ and $\mathcal{N}=2$ Wilson loops starts at three loops is ultimately related, as we saw, to the finiteness of the theory. The reason is the non-renormalization of the quadratic term in the effective action, the same non-renormalization property that guarantees the vanishing of the beta function. At any order of perturbation theory, the Wilson loop vev is given by a combination of rational numbers and zeta-functions whose argument is correlated with the order of perturbation theory. It would be interesting to understand this transcendentality property directly from Feynman diagrams.

At strong coupling we computed the effective string tension, which turns out to depend logarithmically on the 't~Hooft coupling. The most straightforward interpretation of this result is that the curvature of $AdS_5$ in the dual geometry decreases logarithmically with $\lambda $. Such an explanation implicitly assumes that the Wilson loop can be computed semiclassically in string theory. But the dual string theory may not have semiclassical regime\footnote{The discussion of possible quantum effects in the Wilson loop computation on the string side can be found in \cite{Rey:2010ry}.} or may be semiclassical only in some approximate sense, for instance if $AdS_5$ has small curvature at strong coupling, while the compact factor $X^5$ remains highly curved. If this is true, (\ref{stringtension}) is just an effective string tension defined through the parameterization (\ref{WT}) of the Wilson loop vev. In either case, it would be interesting to compute the Wilson loop  in the dual string theory.

Pestun's results are fairly general and potentially apply to any $\mathcal{N}=2$ theory on $S^4$. It would be interesting to repeat the large-$N$ calculation of the Wilson loops for other superconformal, and perhaps also massive theories, although in the latter case the theory on $S^4$ is not equivalent to the theory in flat space. The closest superconformal theory to the one we studied is an interpolating theory with two gauge groups, which connects a $\mathbb{Z}_2$ orbifold of $\mathcal{N}=4$ super-Yang-Mills to $\mathcal{N}=2$ SCYM by a continuous deformation \cite{Gadde:2009dj,Gadde:2010zi,Gadde:2010ku,Pomoni:2011jj,Liendo:2011xb,Rey:2010ry}. At the orbifold point, the Wilson loop vev grows exponentially with $\sqrt{\lambda }$ according to (\ref{N=4}). It would be interesting to investigate the interpolation to  $\mathcal{N}=2$ SCYM, where the scaling is logarithmic. The transition may involve a non-analytic behavior \cite{Rey:2010ry}.

\subsection*{Acknowledgments}
We would like to thank N.~Drukker for participation in the initial stages of this project
and for many illuminating discussions during the course of this work.
We are grateful to J.~Gomis, R.~Janik, V.~Kazakov, R.~Poghossian, S.-J.~Rey, S.~Shatashvili, B.~Stefanski and D.~Young for interesting discussions. K.Z. would like to thank GGI, Florence for hospitality  and INFN for partial support during the course of this work.
The work of K.Z. was supported in part by the Swedish Research Council
under contract 621-2007-4177, in part by the ANF-a grant 09-02-91005,
in part by the RFFI grant 10-02-01315, and in part
by the Ministry of Education and Science of the Russian Federation
under contract 14.740.11.0347.

\appendix
\section{Higher orders at weak coupling}\label{HOweak}

\begin{figure}[t]
\centerline{\includegraphics[width=10cm]{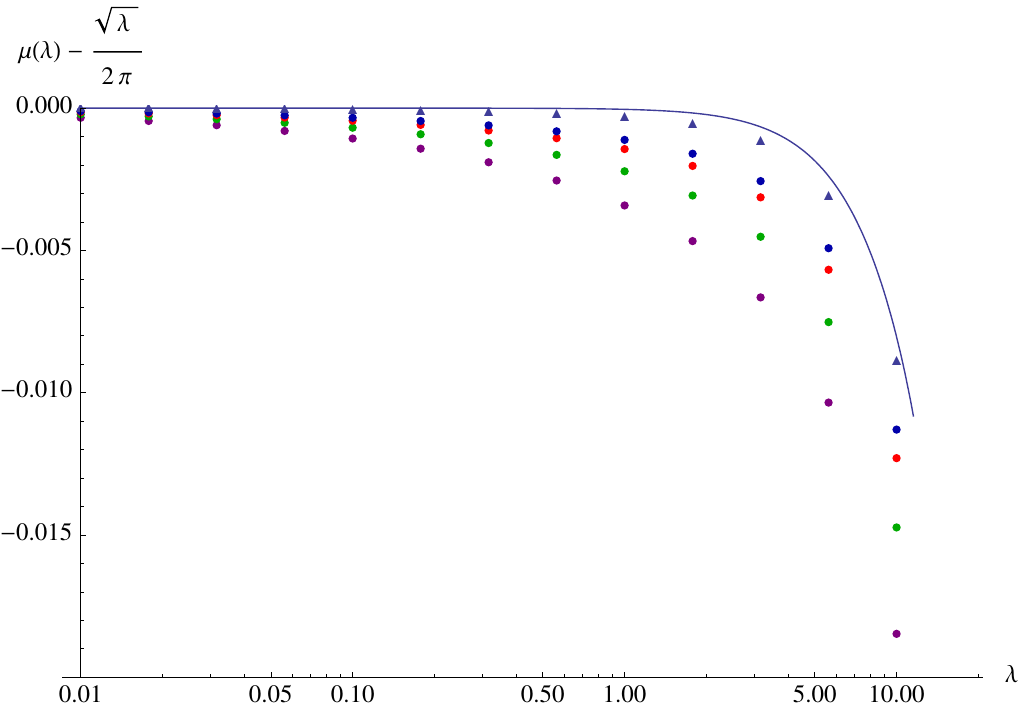}}
\caption{\label{pic:muweak}\small The data denoted as circles are obtained by solving numerically the saddle-point equation with $N=100$ (purple),  $N=200$ (green), $N=400$ (red) and $N=600$ (blue). The triangles are associated to the points extrapolated to $N=\infty$. The  solid line is obtained from the expression (\ref{ExpWeak3}).  }
\end{figure}

Let's approximate  the function $K(z)$ with a truncation of the  Taylor expansion  (\ref{Kseries}), i.e.
\begin{equation}\label{KseriesM}
 K(z)\approx-2\sum_{n=1}^{M}(-1)^n\zeta(2n+1)z^{2n+1}
\end{equation}
where $M$ is a positive integer. Inserting the  (\ref{KseriesM}) in the expression (\ref{Fredholm}),  we obtain
\begin{eqnarray}\label{FredholmM}\nonumber
 \rho (x)&=&\bigg(\frac{8\pi }{\lambda } -\frac{2}{\pi}\sum_{n=1}^{M}(-1)^n\zeta(2n+1)\\&& \times \sum_{k=1}^{n}\binom{2n+1}{2k}\, m_{2k} \sum_{r=0}^{n-k}\frac{(-1)^r}{r!}\,(-1/2)_r\, x^{2(n-k-r)}\mu^{2r}\, \bigg)\sqrt{\mu ^2-x^2}\qquad
\end{eqnarray}
where we defined the $r$-th moment
\begin{equation}
m_r=\int_{-\mu}^{\mu}dz\, \rho (z) z^r=\langle z^r \rangle\, ,
\end{equation}
we  introduced the Pochhammer symbol $(a)_n=a(a-1)\dots (a-n+1)$ and we used
\begin{equation}
\strokedint_{-\mu }^{\mu }\frac{dy}{x-y}\,\,\frac{y^n}{\sqrt{\mu ^2-y^2}} =-\pi\sum_{k=0}^{\left[ \frac{n-1}{2}\right]}\frac{(-1)^k}{k!}\,(-1/2)_k\, x^{n-1-2k}\mu^{2k}\, .
\end{equation}
The normalization condition reads
\begin{eqnarray}\label{NormM}\nonumber
 1&=&\bigg(\frac{4\pi^2\mu^2 }{\lambda } -\sum_{n=1}^{M}(-1)^n\zeta(2n+1) \\&& \times \sum_{k=1}^{n}\binom{2n+1}{2k}\, m_{2k}\, \mu^{2+2(n-k)}\sum_{r=0}^{n-k}\frac{(-1)^r}{r!}\,(-1/2)_r\, \frac{C_{n-k-r}}{2^{2(n-k-r)}}\, \bigg)\qquad
\end{eqnarray}
where $C_n$ is the $n$-th Catalan number.  The $M$ moments $m_{2i}$, $i=1,\dots,M$,  can be  computed using the density (\ref{FredholmM}), giving the following  system of  $M$ equations
\begin{eqnarray}\label{SystemM}\nonumber
 m_{2i}&=&\bigg(\frac{4\pi^2\mu^{2i+2}\, C_i }{2^{2i}\,\lambda  } -\sum_{n=1}^{M}(-1)^n\zeta(2n+1)\\&& \times\sum_{k=1}^{n}\binom{2n+1}{2k}\, m_{2k}\, \mu^{2+2(i+n-k)}\sum_{r=0}^{n-k}\frac{(-1)^r}{r!}\,(-1/2)_r\, \frac{C_{i+n-k-r}}{2^{2(i+n-k-r)}}\, \bigg)\qquad
\end{eqnarray}
where $i=1,\dots,M$. The linear system (\ref{SystemM}) can be be used   to  express the moments $m_{2i}$  in terms of $\mu$ and $\lambda$ and  the normalization condition  (\ref{NormM}) express $\mu$ as a function of $\lambda$.  The approximate  solution (\ref{FredholmM}) therefore can be written  explicitly  at  any  order of  approximation $M$ and can be used to compute the expectation value of the circular Wilson loop (\ref{WLoop}). Expanding $K(z)$ up to $O(z^{2M+1})$, it is possible to compute $\mu(\lambda)$ up to $O(\lambda^{3/2+M})$ and the expectation value of the circular Wilson loop up to $O(\lambda^{M+2})$. For instance, for $M=5$ we obtain
\begin{eqnarray}\label{ExpWeak3}\nonumber
\mu&=&\frac{\sqrt{\lambda }}{2 \pi }-\frac{3 \zeta (3) \lambda ^{5/2}}{256 \pi ^5}+\frac{5 \zeta (5) \lambda ^{7/2}}{512 \pi
   ^7}+\frac{7 \left(9 \zeta (3)^2-65 \zeta (7)\right) \lambda ^{9/2}}{65536 \pi ^9}\\
   &+&\frac{3 (861 \zeta (9)-340 \zeta (3) \zeta (5)) \lambda ^{11/2}}{524288
   \pi ^{11}}\nonumber \\ 
   &+&\frac{\left(-891 \zeta (3)^3+7900 \zeta (5)^2+13965 \zeta (3) \zeta
   (7)-30261 \zeta (11)\right) \lambda ^{13/2}}{8388608 \pi ^{13}}+\ldots \nonumber  \\
\end{eqnarray}
and
\begin{eqnarray}\label{WLWeak3}\nonumber
 W(C_{\rm circle})&=&
1+\frac{\lambda }{8}+\frac{\lambda ^2}{192}+\left(\frac{1}{9216}-\frac{3 \zeta (3)}{512 \pi
   ^4}\right) \lambda ^3\\\nonumber &+&\left(\frac{1}{737280}-\frac{2 \pi ^2 \zeta (3)-15 \zeta (5)}{4096 \pi ^6}\right) \lambda ^4\\\nonumber
  &+&\left(\frac{1}{88473600}-\frac{3 \pi ^4 \zeta (3)-65 \pi ^2 \zeta (5)-12 \left(9 \zeta (3)^2-35 \zeta (7)\right)}{196608 \pi ^8}\right) \lambda^5
\nonumber \\
&+&\bigg(\frac{1}{14863564800} +\frac{-2 \pi ^2 \zeta (3)+85  \zeta (5)}{7864320 \pi ^{6}}\nonumber \\&&+\frac{  \pi^2\left(180
   \zeta (3)^2-637 \zeta (7)\right)-45 (60 \zeta (3) \zeta (5)-91 \zeta
   (9))}{3145728 \pi ^{10}}\bigg) \lambda ^6 \nonumber\\
   &+&\bigg(\frac{1}{3329438515200}+\frac{- \pi ^2 \zeta (3)+70 \zeta (5)}{377487360 \pi
   ^{6}}\nonumber\\&&+\frac{3 \pi ^2 \left(108 \zeta (3)^2-343 \zeta (7)\right)-126  (110 \zeta (3) \zeta
   (5)-153 \zeta (9))
 }{150994944 \pi
   ^{10}}\nonumber\\&&-\frac{27 \left(360 \zeta (3)^3-1900 \zeta (5)^2-3360 \zeta (3) \zeta (7)+4697 \zeta (11)\right)}{150994944 \pi
   ^{12}}\bigg) \lambda ^7\nonumber 
\\&+&O(\lambda ^8)
\end{eqnarray}
The expression (\ref{ExpWeak3}) is compared to the numerical data in in fig.~\ref{pic:muweak}. 

\section{Numerical results for the endpoint position at strong coupling}
\label{app:endpoint}
 
We have shown in section \ref{sec:strong} that in the regime of   large $\lambda$, it results $C=\lambda\, {\rm e}\,^{-\pi \mu /2}/\sqrt{\mu }$ where $C$ is a constant estimated as $C\approx 14.60$.  In order to check this result  we have performed an extensive numerical analysis. In doing so we face two technical difficulties. First, it is difficult to calculate the position of the endpoint numerically with good precision, because the density goes to zero there and at a finite $N$ there will be few datapoints close to $x=\mu $, which is visible in fig.~\ref{pic:rho001}, \ref{pic:rhoinf} and \ref{pic:edge}. We can estimate how large $N$ should be. The density in the interval of interest (region III) scales as $\rho \sim \sqrt{\mu }/\lambda $ (eq.~(\ref{III})) and changes on the distances of order one. The distance between adjacent eigenvalues is thus of order $\Delta x\sim 1/N\rho \sim \lambda /N\sqrt{\mu }$. This should be at least as small as the scale of variation of the density. We thus need $N\gtrsim \lambda /\sqrt{\mu }\sim \lambda (\ln\lambda )^{-1/2}$ to reach reasonable numerical accuracy. But  the calculational cost grows as $N^2$ and it is impractical to go beyond $N\sim 1000$ by running {\it Mathematica} on a laptop. We cannot thus reach very big $\lambda $, and very big $N$, but on the other hand to estimate $C$ we need to know $\mu $ with a precision that grows exponentially with $\mu $. Our strategy was to compute $\mu $ for several values of $N$ and then numerically extrapolate to $N=\infty $, and subsequently to extrapolate in $\lambda $.

\begin{figure}[t]
\centerline{\includegraphics[width=10cm]{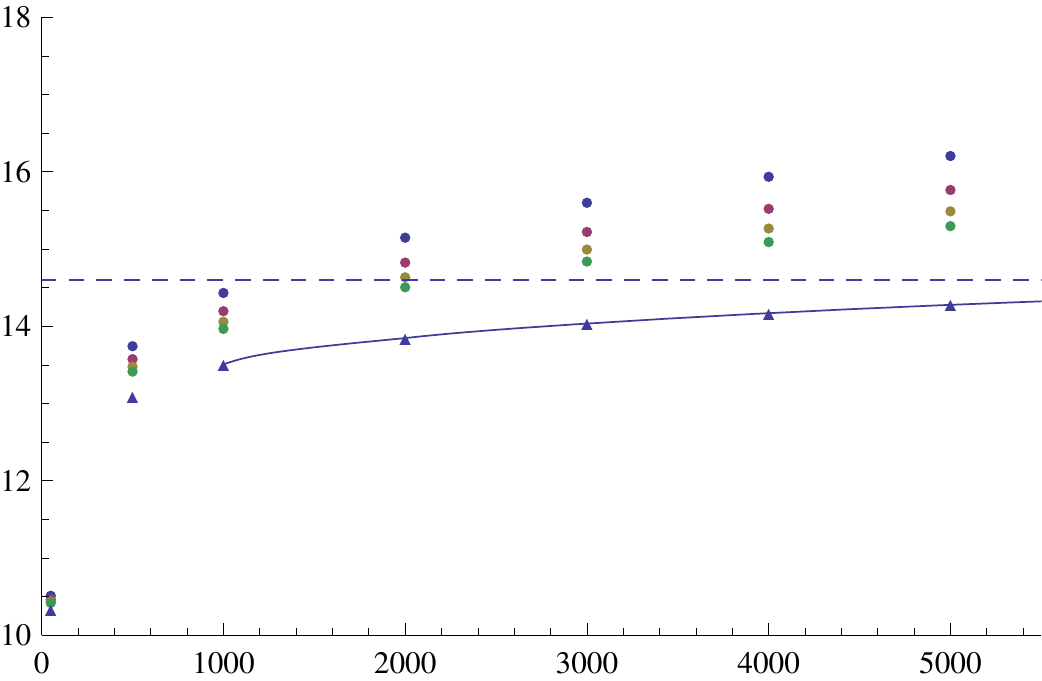}}
\caption{\label{pic:cplot}\small The data denoted as circles are obtained by solving numerically the saddle-point equation with $N=400$ (purple),  $N=600$ (blue), $N=800$ (red) and $N=1000$ (green). The triangles are associated to the points extrapolated to $N=\infty$. The  solid line is the interpolating function for the $N=\infty$ points and the dashed line is the analytical expectation $C=14.60$.  }
\end{figure}     

We have numerically solved the  saddle-point equation in the range $50<\lambda<5000$ (as explained above we cannot get reliable results for larger $\lambda $) considering the  number of eigenvalues $N=400$, $N=600$, $N=800$ and $N=1000$ and computed the associated  endpoints $\mu$, see fig.~\ref{pic:cplot}.  These data have been used to extrapolate the value of the endpoint $\mu$ to $N=\infty$ for any given $\lambda$.  We have used the values of $\mu $ at $N=\infty$ to construct an interpolating function   $C_\text{INT}(\lambda)=\lambda\, {\rm e}\,^{-\pi \mu /2}/\sqrt{\mu }$ such that $C_\text{INT}(\infty)$ gives the value of the  constant $C$. We obtain that $C=C_\text{INT}(\infty)\approx 15.02$, in agreement with the analytical result, certainly within errorbars of both calculations.
   
\section{Derivation of Wiener-Hopf solution}\label{app:wienerhopf}

The generalized Wiener-Hopf method replaces the Fourier transform when  an integral equation with a difference kernel is defined on a finite interval. We follow the variant of the method described in the appendix of \cite{Japaridze:1984dz}.

By Fourier transforming eq.~(\ref{ineq}), we get:
\begin{equation}\label{inFourier}
 \int_{-\infty }^{+\infty }\frac{d\omega }{2\pi }\,\,
 \,{\rm e}\,^{-i\omega x}\left(
 \frac{\cosh\omega }{2\sinh^2\frac{\omega }{2}}\,\rho (\omega )
 -\frac{1}{2\sinh^2\frac{\omega }{2}}-F(\omega )
 \right)=0,
\end{equation}
where $F(\omega )$ is the Fourier transform of the right-hand side.

If (\ref{inFourier}) were to hold for any $x$, we would immediately find  $\rho (\omega )$, but the equation holds only for $x\in(-\mu ,\mu )$, and the integrand need not be equal to zero. The remainder can be represented as a sum of two rapidly oscillating functions:
\begin{equation}\label{intermed}
 \frac{\cosh\omega }{2\sinh^2\frac{\omega }{2}}\,\rho (\omega )
 =\frac{1}{2\sinh^2\frac{\omega }{2}}+F(\omega )+
 \,{\rm e}\,^{-i\mu \omega }X_-(\omega )+\,{\rm e}\,^{i\mu \omega }X_+(\omega ),
\end{equation}
where $X_\pm(\omega )$ does not contain negative/positive frequencies and con\-se\-quent\-ly can be analytically continued to the upper/lower half-plane of complex $\omega $.

To find $\rho (\omega )$ we can use the decomposition (\ref{GtoG+G-}) and
projection operators onto the positive and negative frequency parts. For any function, the projections are defined by an integral transform:
\begin{equation}
 \mathcal{F}(\omega )=\mathcal{F}_+(\omega )+\mathcal{F}_-(\omega ),\qquad
 \mathcal{F}_\pm(\omega )=\pm\int_{-\infty }^{+\infty }\frac{d\omega '}{2\pi i}\,\,
 \frac{\mathcal{F}(\omega ')}{\omega '-\omega \mp i0}\,.
\end{equation}
The functions $\mathcal{F}_\pm(\omega )$ are analytic in the upper/lower half-planes. 

Multiplying both sides of  (\ref{intermed}) by $G_+(\omega )\,{\rm e}\,^{-i\mu \omega }$, and taking the negative-frequency part, we find:
\begin{eqnarray}\label{trm}
 \frac{\rho (\omega )\,{\rm e}\,^{-i\mu \omega }}{G_-(\omega )}
 &=&
 \left[G_+(\omega )\left(
 \frac{1}{2\sinh^2\frac{\omega }{2}}+
 F(\omega )\right)\,{\rm e}\,^{-i\mu \omega }\right]_-
\nonumber \\
 &&+\left[G_+(\omega )X_-(\omega )\,{\rm e}\,^{-2i\mu \omega }\right]_-.
\end{eqnarray}
The functions $\rho (\omega )\,{\rm e}\,^{-i\mu \omega }$ and $F(\omega )\,{\rm e}\,^{-i\mu \omega }$ are analytic in the lower half-plane, because their Fourier images vanish for positive $x$. For this reason it was possible to omit the negative-frequency projection on the left-hand side. The function $1/\sinh^2(\omega /2)$ has poles in the lower half-plane, but those poles are cancelled by zeros of $G_+(\omega )$.

Although resulting equation still contains an unknown function $X_-(\omega )$, one can  argue that the last term in (\ref{trm}) is exponentially small in $\mu $. To the first approximation we can neglect it. The density then is given by the standard solution of the Wiener-Hopf problem on a semi-infinite interval:
\begin{equation}\label{WH1}
 \rho (\omega )=G_-(\omega )\,{\rm e}\,^{i\mu \omega }
 \left[G_+(\omega )\left(
 \frac{1}{2\sinh^2\frac{\omega }{2}}+
 F(\omega )\right)\,{\rm e}\,^{-i\mu \omega }\right]_-
 .
\end{equation}
As we have already mentioned this gives an accurate description of the density at $x>0$, and in particular satisfies the correct boundary condition at $x=\mu $: $\rho (x>\mu )=0$, since $\rho (\omega )\,{\rm e}\,^{-i\mu \omega }$ is manifestly analytic in lower half-plane. But the boundary condition at $x=-\mu $ is not satisfied and we thus cannot trust this approximate solution at $x<0$.

The solution in the main text is obtained by closing the contour of integration in
 the negative-frequency projection   in the lower half-plane
  and picking up the poles of $G_+(\omega )$. 
\section{Second iteration at strong coupling}\label{1interation}

Here we compute the correction to the strong coupling solution due to the second line in (\ref{ineq}):
\begin{eqnarray}
 \delta F(x)&=& -\frac{1}{\pi }\int_{|w|>\mu }^{}\frac{dw}{w-x}\,\,\sqrt{
 \frac{\mu ^2-x^2}{w^2-\mu ^2}}\int_{}^{}dy\,\rho (y)\left[
 \left(w-y\right)\coth\pi \left(w-y\right)
 \right.
\nonumber \\
&&\left.
 -w\coth\pi w
 \right].
\end{eqnarray}
Assuming that $x\sim \mu $ and $\mu -x\sim \mu $ (this corresponds to the region II), and taking into account that the $w $ integral is dominated by a small neighborhood of the endpoints $|w\pm \mu |\sim 1$, we can  replace $1/(w-x)$ by $1/(\pm \mu- x)$. The whole expression then considerably simplifies:
\begin{equation}
 \delta F(x)\approx -\frac{2A\mu }{\pi \sqrt{\mu ^2-x^2}}\,,
\end{equation}
where
\begin{equation}
 A=\int_{\mu }^{\infty }\frac{dw}{\sqrt{w^2-\mu ^2}}\,
 \int_{}^{}dy\,\rho (y)\left[
 \left(w-y\right)\coth\pi \left(w-y\right)
  -w\coth\pi w
 \right].
\end{equation}

Upon the Fourier transform, $A$ becomes
\begin{equation}
 A=\frac{\pi }{2}\int_{-\infty }^{+\infty }\frac{d\omega }{2\pi i}\,\,
 \frac{1-\rho (\omega )}{2\sinh^2\frac{\omega }{2}}\,H^{(2)}_0(\mu \omega ),
\end{equation}
where $H^{(2)}_0(x )$ is the Hankel function of the second kind. The integral can be calculated by closing the contour of integration into the lower half-plane and picking the poles at the zeros of $\sinh(\omega /2)$:
\begin{equation}
 A=2\sum_{n=1}^{\infty }\left.\frac{d}{d\kappa }\left[\left(1-\rho (-i\kappa)\right)K_0(\mu \kappa )\right]\right|_{\kappa =2\pi n}.
\end{equation}
Using the leading-order solution in the form (\ref{densityfin1}) we find:
\begin{eqnarray}
 A&=&2i\sum_{n=1}^{\infty }\left.\frac{d}{d\kappa }\left[\,{\rm e}\,^{\mu \kappa }K_0(\mu \kappa )
 \vphantom{\sum_{m=1}^{\infty }\frac{r_m\,{\rm e}\,^{-\mu \nu _m}F(-i\nu _n)\left(\nu _m-\nu _0\right)\left(\kappa +\nu _0\right)}{\left(\nu _m+\nu _0\right)\left(\kappa -\nu _0\right)\left(\kappa -\nu _m\right)}}
 G_-(-i\kappa )\right.\right.
\nonumber \\
&&\left.\left.\times 
 \sum_{m=1}^{\infty }\frac{r_m\,{\rm e}\,^{-\mu \nu _m}F(-i\nu _m)\left(\nu _m-\nu _0\right)\left(\kappa +\nu _0\right)}{\left(\nu _m+\nu _0\right)\left(\kappa -\nu _0\right)\left(\kappa -\nu _m\right)}
 \right]\right|_{\kappa =2\pi n}.
\end{eqnarray}
At large-$\mu $,
\begin{equation}
 K_0(\mu \kappa )\simeq \,{\rm e}\,^{-\mu \kappa }\sqrt{\frac{\pi }{2\mu \kappa }}
\end{equation}
and using the similar approximation  (\ref{largemu}) for $F(-i\nu _m)$, we get:
\begin{equation}
 A=\frac{a}{\lambda }\,
\end{equation}
where $a$ is a numerical constant:
\begin{eqnarray}
 a&=&-2^{7/2}\pi ^{-3/2}\sum_{m,n=1}^{\infty }
 \frac{mr_m}{\left(2m+1\right)^{3/2}\left(m+1\right)}
\nonumber \\
&&\times
 \frac{d}{dn}\left[
 \frac{\Gamma \left(2n+\frac{3}{2}\right)}{2^{2n}n^{3/2}\,\Gamma ^2(n)\left(2n-\frac{1}{2}\right)\left(2n-m-\frac{1}{2}\right)}
 \right]=1.0232.
\end{eqnarray}

This gives a correction to $F(\omega )$:
\begin{equation}
 \delta F(\omega )=-\frac{2a\mu J_0(\mu \omega )}{\lambda }
\end{equation}
and
\begin{equation}\label{corr}
 \delta F(-i\nu _n)=-\frac{2a\mu I_0(\mu \nu _n )}{\lambda }\approx
 -a\sqrt{\frac{2}{\pi \nu _n}}
 \,\frac{\sqrt{\mu }\,{\rm e}\,^{\mu \nu _n}}{\lambda }\,.
\end{equation}
Substituting this into (\ref{sums}) we find the correction to the normalization condition  of the form (\ref{mularge}) with the constant $C$ thus shifted by
\begin{equation}
 \delta C=-\frac{2a}{\pi r_0}\,\sum_{n=0}^{\infty }\frac{r_n}{\left(2n+1\right)^{1/2}\left(n+1\right)}=-0.56.
\end{equation}
There is no parametric suppression with respect to the leading order, but numerically the corrections is rather small.

We can also calculate the correction to the coefficient $R$ in the Wilson loop vev (\ref{vevscaling}):
\begin{equation}\label{corrtoR}
 \delta R=-\frac{5a}{4\pi }\,\sum_{n=1}^{\infty }\frac{nr_n}{\left(n+\frac{1}{2}\right)^{1/2}\left(n+1\right)\left(3-2n\right)}=-0.37.
\end{equation}

\end{document}